\documentclass[nofootinbib,twocolumn,aps,superscriptaddress,preprintnumbers]{myRevtex4}
\pdfoutput=1

\usepackage{amsmath,amssymb,amsfonts} 
\usepackage{graphicx}
\usepackage{subfigure}
\usepackage{color}
\usepackage{xspace}
\usepackage{natbib}   
\usepackage{tabularx}
\usepackage{hyperref}

\definecolor{darkblue1}{rgb}{0,0,.4}
\hypersetup{breaklinks=true, 
            colorlinks=true, 
            linkcolor=darkblue1, 
            menucolor=darkblue1, 
            urlcolor=darkblue1,
            citecolor=darkblue1
}

\bibstyle{plain}
\mathchardef\Upsilon="7107
\def\Y#1S{\ensuremath{\Upsilon{(#1S)}}\xspace}

\newcommand{\nlojet}{NLOJET++\xspace} 
\newcommand{\AKT}{anti-${k_t}$\xspace}
\newcommand{\applgrid}{APPLGRID\xspace}
\newcommand{\pythia}{PYTHIA\xspace} 
\newcommand{\fastjet}{FASTJET\xspace}  
\newcommand{\herwigpp}{HERWIG\raisebox{0.1ex}{++}\xspace}

\newcommand{\mZ}{\ensuremath{M_Z}\xspace}
\newcommand{\as}{\ensuremath{\alpha_{\scriptscriptstyle S}}\xspace}
\newcommand{\asNomVec}{\ensuremath{\bar{\alpha}_{\scriptscriptstyle S}^{\rm Nom}}\xspace}
\newcommand{\asNomi}{\ensuremath{\alpha_{\scriptscriptstyle S}^{\rm Nom,i}}\xspace}
\newcommand{\asAvVec}{\ensuremath{\bar{\alpha}_{\scriptscriptstyle S}^{\rm Av}}\xspace}
\newcommand{\asAv}{\ensuremath{\alpha_{\scriptscriptstyle S}^{\rm Av}}\xspace}

\newcommand{\XsecNomVec}{\ensuremath{\bar{\sigma}_{\scriptscriptstyle S}^{\rm Nom}}\xspace}
\newcommand{\XsecThVec}{\ensuremath{\bar{\sigma}_{\scriptscriptstyle S}^{\rm Th}}\xspace}
\newcommand{\chiSq}{\ensuremath{\chi^2}\xspace}

\newcommand{\asZ}{\ensuremath{\as(\mZ^2)}\xspace}

\newcommand{\Kbar    }{\kern 0.2em\overline{\kern -0.2em K}{}\xspace}
\newcommand{\Dbar    }{\kern 0.2em\overline{\kern -0.2em D}{}\xspace}

\newcommand{\Kz      }{\ensuremath{K^0}\xspace}
\newcommand{\Kzb     }{\ensuremath{\Kbar^0}\xspace}
\newcommand{\KzKzb   }{\ensuremath{\Kz \kern -0.16em \Kzb}\xspace}
\newcommand{\Kp      }{\ensuremath{K^+}\xspace}
\newcommand{\Km      }{\ensuremath{K^-}\xspace}

\newcommand{\KpKm    }{\ensuremath{\Kp \kern -0.16em \Km}\xspace}

\newcommand{\tev}{\ensuremath{\mathrm{\,Te\kern -0.1em V}}\xspace}
\newcommand{\gev}{\ensuremath{\mathrm{\,Ge\kern -0.1em V}}\xspace}
\newcommand{\mev}{\ensuremath{\mathrm{\,Me\kern -0.1em V}}\xspace}
\newcommand{\kev}{\ensuremath{\mathrm{\,ke\kern -0.1em V}}\xspace}
\newcommand{\ev}{\ensuremath{\mathrm{\,e\kern -0.1em V}}\xspace}
\newcommand{\gevc}{\ensuremath{{\mathrm{\,Ge\kern -0.1em V\!/}c}}\xspace}
\newcommand{\mevc}{\ensuremath{{\mathrm{\,Me\kern -0.1em V\!/}c}}\xspace}
\newcommand{\gevcc}{\ensuremath{{\mathrm{\,Ge\kern -0.1em V\!/}c^2}}\xspace}
\newcommand{\mevcc}{\ensuremath{{\mathrm{\,Me\kern -0.1em V\!/}c^2}}\xspace}

\newcommand{\bei}{\begin{itemize}}
\newcommand{\eei}{\end{itemize}}
\newcommand{\ben}{\begin{enumerate}}
\newcommand{\een}{\end{enumerate}}
\newcommand{\beq}{\begin{equation}}
\newcommand{\eeq}{\end{equation}}
\newcommand{\beqn}{\begin{eqnarray}}
\newcommand{\eeqn}{\end{eqnarray}}
\newcommand{\beqns}{\begin{eqnarray*}}
\newcommand{\eeqns}{\end{eqnarray*}}

\renewcommand{\arraystretch}{1.25}

\newcommand{\pt}{\ensuremath{p_{\rm T}}\xspace}
\newcommand{\ptmin}{\ensuremath{p_{\rm T}^{\rm min}}\xspace}
\newcommand{\ptmax}{\ensuremath{p_{\rm T}^{\rm max}}\xspace}
\newcommand{\absy}{\ensuremath{|y|}\xspace}
\def\@citex[#1]#2{\if@filesw\immediate\write\@auxout{\string\citation{#2}}\fi
  \@tempcnta\z@\@tempcntb\m@ne\def\@citea{}\@cite{\@for\@citeb:=#2\do
    {\@ifundefined
       {b@\@citeb}{\@citeo\@tempcntb\m@ne\@citea
        \def\@citea{,\penalty\@m\ }{\bf ?}\@warning
       {Citation `\@citeb' on page \thepage \space undefined}}%
    {\setbox\z@\hbox{\global\@tempcntc0\csname b@\@citeb\endcsname\relax}%
     \ifnum\@tempcntc=\z@ \@citeo\@tempcntb\m@ne
       \@citea\def\@citea{,\penalty\@m}
       \hbox{\csname b@\@citeb\endcsname}%
     \else
      \advance\@tempcntb\@ne
      \ifnum\@tempcntb=\@tempcntc
      \else\advance\@tempcntb\m@ne\@citeo
      \@tempcnta\@tempcntc\@tempcntb\@tempcntc\fi\fi}}\@citeo}{#1}}

\def\@citeo{\ifnum\@tempcnta>\@tempcntb\else\@citea
  \def\@citea{,\penalty\@m}%
  \ifnum\@tempcnta=\@tempcntb\the\@tempcnta\else
   {\advance\@tempcnta\@ne\ifnum\@tempcnta=\@tempcntb \else
\def\@citea{--}\fi
    \advance\@tempcnta\m@ne\the\@tempcnta\@citea\the\@tempcntb}\fi\fi}
\newenvironment{myquote}
               {\list{}{\leftmargin0cm\indent}%
                \item\relax}
               {\endlist}
\newcommand\allFontSize{\footnotesize}
\newcommand\detailsSize{\allFontSize}
\newenvironment{details}%
{\begin{myquote}\detailsSize}{\end{myquote}}

\begin{document}

\preprint{\vbox{\hbox{CERN-PH-EP-2012-064, DESY-12-039}}}

\title{\boldmath Evaluation of the Strong Coupling Constant \as Using the ATLAS Inclusive Jet Cross-Section Data}

\author{B.~Malaescu}
\affiliation{CERN, CH--1211, Geneva 23, Switzerland}
\author{P.~Starovoitov}
\affiliation{DESY, 22607, Notkestra\ss{e} 85, Hamburg, Germany}

\date{\today}

\begin{abstract}
We perform a determination of the strong coupling constant using the latest ATLAS inclusive jet cross section data, from proton-proton collisions at $\sqrt{s}=7{\rm TeV}$,
and their full information on the bin-to-bin correlations.
Several procedures for combining the statistical information from the different data inputs are studied and compared.
The theoretical prediction is obtained using NLO QCD, and it also includes non-perturbative corrections.
Our determination uses inputs with transverse momenta between 45 and 600~GeV, the running of the strong coupling being also tested in this range.
Good agreement is observed when comparing our result with the world average at the Z-boson scale, as well as with the most recent results from the Tevatron.
\end{abstract}

\maketitle

\section{Introduction}

The strong coupling constant \as is one of the fundamental parameters of the Standard Model of particle physics~(SM).
Testing the energy dependence~(running) of \as over a wide range provides an implicit test of QCD and probes potential effects from ``New Physics''.

In this paper we present a determination of \as exploiting the recently published ATLAS inclusive jet cross section data~\cite{Aad:2011fc},
allowing for the first test of the running of \as up to the TeV scale.
The only other input based on experimental measurements consists of the NLO QCD proton parton density functions~(PDFs), as obtained by various groups
from other independent data sets.
These PDFs allow us to compute, for a given \as value, the perturbative NLO QCD prediction of the inclusive jet cross section, which is then corrected for non-perturbative effects.
Our evaluation procedure goes through the determination of \asZ using the measured cross section in each data bin, followed by the propagation
of the experimental uncertainties and bin-to-bin correlations.
Finally, an unbiased averaging procedure is used and the uncertainties are propagated to the corresponding result.

In Sec.~\ref{Sec:InputData} we discuss the information provided by the ATLAS inclusive jet cross section data,
while in Sec.~\ref{Sec:Theory} we describe the theoretical prediction used in the \as determination.
Several possible procedures of data treatment are discussed in Sec.~\ref{Sec:EvaluationProcedure}, while the corresponding results are presented in
Sec.~\ref{Sec:Results}.
Our conclusions from this study are discussed in Sec.~\ref{Sec:Conclusions}.

\section{Input Data}
\label{Sec:InputData}

In this study we exploit the unfolded double-differential ATLAS inclusive jet cross section data, in bins of the absolute rapidity 
$(\absy)$ and as a function of the jet transverse momentum~(\pt), together with the statistical and systematic uncertainties, and their correlations~\cite{Aad:2011fc}.
Jets are defined using the \AKT jet algorithm~\cite{Cacciari:2008gp} implemented in the \fastjet~\cite{Cacciari:2005hq} package.
This measurement, performed for \AKT jets with a distance parameter R=0.4 and R=0.6 respectively, covers a large phase-space region going up to 4.4 in absolute rapidity and from 20~GeV 
to more than 1~TeV in \pt.
The detailed analysis of the uncertainties and their correlations is one of the important achievements of this measurement.

The dominant uncertainty in this measurement is due to the jet energy scale~(JES) calibration.
Other smaller systematic uncertainties are due to the luminosity measurement, multiple proton-proton interactions~(``pile-up''), trigger efficiency, jet reconstruction and 
identification, jet energy resolution and deconvolution of detector effects~(unfolding).
The information on the (anti-)~correlations between the various \absy and \pt bins is provided for these components, using a set of 87 independent nuisance parameters.
These nuisance parameters also provide the information on the asymmetries of these uncertainties~(mainly due to the steeply falling \pt spectrum).
This type of information cannot be incorporated in a covariance matrix.
Depending on the rapidity bin, the systematic uncertainties account for $20-60\%$ at low \pt, and $20-40\%$ at high \pt, while they are smaller for
intermediate jet transverse momentum values.

The statistical uncertainties and their (anti-)~correlations are provided through a covariance matrix for each rapidity bin.
Here the correlations are due to the transfers of jets between the different \pt bins~(performed in the unfolding), as well as to the fact
that several jets going into the spectrum can be produced in the same event.
The statistical uncertainties are in general much smaller than the systematics, except for the high \pt region, where they can be larger.

It has been pointed out in~\cite{Aad:2011fc} that, when performing the comparison between data and the theoretical prediction, somewhat larger differences are observed for R=0.4 than for R=0.6 jets.
However, the $\chi^{2}/{\rm dof}$, computed taking into account the correlations, is reasonable in both cases.
In our study, we do not perform a simultaneous treatment of the two jet sizes, because the correlations of the uncertainties of these measurements~(expected to be large) are not available.
We use the measurement of jets with R=0.6 for determining our nominal result, while the comparison with the result using jets with R=0.4 provides a systematic uncertainty.

\section{Theoretical Prediction}
\label{Sec:Theory}

Theoretical predictions for the inclusive jet cross sections are calculated in perturbative~QCD at next-to-leading order~(NLO)  using the \nlojet~\cite{Nagy:2003tz}
program, while the effects of hadronisation and underlying event activity on jet production have been studied using Monte Carlo generators. 
Each bin of the perturbative cross section is multiplied by the non-perturbative corrections to obtain the prediction at the particle level that can be compared to the data.

 \subsection{Perturbative QCD calculations}

The inclusive jet cross section is calculated using the CT10~\cite{Lai:2010vv} NLO parton density functions.
The $\pt$ and rapidity binnings follow the one from ATLAS inclusive jet cross section measurement~\cite{Aad:2011fc}.
For the sake of simplicity, the renormalisation $\mu_{\rm R}$ and the factorisation $\mu_{\rm F}$ scales for the central result are set to be the same in the calculation 
\begin{equation*}
  \mu = \mu_{\rm R} = \mu_{\rm F} = \pt^{\rm max},
  \label{eq:ptscale}
\end{equation*}
where $\pt^{\rm max}$ is the transverse momentum of the leading jet in a given rapidity bin.

To allow for fast and flexible evaluation of the cross section using different values of $\as$, as well as for evaluation of the PDF and scale uncertainties,
the \applgrid~\cite{Carli:2010rw} software was interfaced with \nlojet in order to calculate the perturbative coefficients once and store them in a look-up table.

 \subsection{PDF inputs}

There are several PDF fitter groups who provide alternative PDF sets using slightly different theoretical approaches and various experimental data from different scattering processes as input. Usually PDF groups provide the central (best-fit, given an \asZ value) PDF set together with a collection of  uncertainty PDF sets. The central set is used for the theory prediction, while the uncertainty sets are needed for the evaluation of the PDF uncertainties in the calculation. 

The best-fit CT10~PDF set is obtained with \asZ$=0.118$. In order to facilitate the studies of the dependence of observables upon the strong coupling CT10~PDF  is provided with the special list of PDF sets, obtained by scanning the \asZ.
The PDF sets in this list were refitted while the \asZ was varied  in the $[0.113, 0.123]$ range in steps of $0.001$. 

Given the $\as$-scan, we have calculated the inclusive jet cross section $\sigma^{Th}(\as, \pt, |y|)$ for each $\as$-point in the scan.
Such a function provides a one-to-one correspondence between the $\asZ$ and the jet cross section values in each $(\pt, |y|)$ bin. For the values of $\as$ outside the scan range (between scan points) a linear extrapolation~(interpolation) is used. 

In contrast to CT10, the central MSTW~2008 fit~\cite{Martin:2009iq} suggests that $\asZ=0.1202$ and MSTW group provides the $\as$-scan in the range $[0.110, 0.130]$.
For the NNPDF~2.1~\cite{Ball:2011mu, Ball:2011uy} the central PDF set is fitted with \asZ fixed to $0.1190$ and the scan is performed for $\as$ values from $0.110$ to $0.130$ .
Finally, for the HERAPDF~1.5~\cite{HERAPDF15} an $\as$ scan has been performed, from $0.114 $ to $0.122$, with the best-fit value at $\asZ=0.1176$. All the PDF sets provide the $\as$-scan with the same stepping.
The nominal \asZ values for the various PDF sets, as well as the ranges of the scans, are summarized in table~\ref{Tab:PDFasRange}.

\begin{table}[t]
 \caption[.]{ \label{Tab:PDFasRange}
   The range of $\as$-scan and the best-fit \asZ\ values for PDF sets used in the analysis.
}
\setlength{\tabcolsep}{0.0pc}
\begin{tabularx}{\columnwidth}{@{\extracolsep{\fill}}ccccc}
\hline\noalign{\smallskip}
PDF set & $\asZ^{\rm best}$ & $\asZ^{\rm min}$ & $\asZ^{\rm max}$ & step \\
\noalign{\smallskip}\hline\noalign{\smallskip}
CT10 & 0.118 & 0.113 & 0.123 & 0.001 \\
MSTW 2008 & 0.1202 & 0.110 & 0.130 & 0.001 \\
HERAPDF 1.5 & 0.1176 & 0.114 & 0.122 & 0.001 \\
NNPDF 2.1 & 0.1190 & 0.110 & 0.130 & 0.001 \\
\noalign{\smallskip}\hline
\end{tabularx}
\end{table}


 \subsection{Non-perturbative corrections}

The fixed-order NLO~QCD calculations predict hard parton-level cross sections, which need to be corrected for non-perturbative effects when compared to data. 
After the hard scattering phase of pp-collision partons emit radiation and evolve down to lower scales. At a very low scale, when strong coupling becomes very large,
the non-perturbative phase of collision starts, where coloured partons form colourless  real hadrons.
This process is called hadronisation.
There is also underlying event~(UE) activity, which happens in parallel to the hard scattering.
The UE consists of interactions of beam remnants as well as from multiple parton interactions.

The hadronisation process leads to a decrease of the jet $\pt$, while the UE adds extra particles to the jet and results in an increase of jet transverse momentum.
The effects of both hadronisation and UE are most pronounced at low $\pt$. Since the effect of UE on the jet $\pt$  increases with increasing size of the jet (jet radius
$R$) while the effect of the hadronisation process decreases, if the jet radius grows, two separate sets of corrections and uncertainties were derived for jets with $R=0.4$ and
$R=0.6$.
The $R$ dependence of the contributions of non-perturbative effects to the jet transverse momentum was calculated in \cite{Dasgupta:2007wa,Dasgupta:2009tm}.
It has been shown that the hadronisation contribution is proportional to $-1/R$, while the UE contribution varies as $R^2$.
The leading $R$ dependence was found to be quite similar for all jet algorithms.
For $R=0.4$, the correction factors are dominated by the effect of hadronisation and are approximately 0.95 at jet $\pt=20$~GeV, increasing closer to unity at higher $\pt$.
For $R=0.6$ the correction factors are dominated by the UE and are approximately 1.6 at jet $\pt=20$~GeV decreasing to between $1.0-1.1$ for jets above $\pt=100$~GeV.

The non-perturbative corrections are evaluated using leading-logarithmic parton shower generators. 
The corrections are derived using \pythia~6.425 with the ATLAS~AUET2B~CTEQ6L1 tune~\cite{ATL-PHYS-PUB-2011-009}.
These are evaluated as the bin-wise ratio of cross section calculated after hadronisation and with simulation of the underlying event activity, to the one at parton-level and without underlying event.
The parton-level cross section in each bin is then multiplied by the corresponding correction:
\beqn
\frac{d^2\sigma^{particle}}{dp_{T}^{jet} d|y|} = \frac{d^2\sigma^{NLO~QCD}}{dp_{T}^{jet} d|y|} \times \delta^{non-pert}\left( p_{T}^{jet}, |y|\right) .
\eeqn

The uncertainty of the non-perturbative corrections is estimated as the maximum spread of the correction factors obtained from \pythia~6.425 using the AUET2B~LO$^{**}$, AUET2~LO$^{**}$, AMBT2B~CTEQ6L1, AMBT1, Perugia~2010, and Perugia~2011 tunes (PYTUNE\_350), and the \pythia~8.150 tune~$4C$~\cite{ATL-PHYS-PUB-2011-008,ATL-PHYS-PUB-2011-009,Skands:2009zm,Skands:2010ak}, as well as the corrections obtained from \herwigpp~2.5.1~\cite{Bahr:2008pv} using the UE7000-2~\cite{ATL-PHYS-PUB-2011-009} tune~(see table~\ref{Tab:tunes}). In our analysis we have used the non-perturbative corrections  evaluated in~\cite{Aad:2011fc}.
  
\begin{table}[t]
 \caption[.]{ \label{Tab:tunes}
   Set of tunes used for the different generators.
}
\setlength{\tabcolsep}{0.0pc}
\begin{tabularx}{\columnwidth}{@{\extracolsep{\fill}}ll}
\hline\noalign{\smallskip}
Generator & Tune\\
\noalign{\smallskip}\hline\noalign{\smallskip}
\pythia~6.425 &  AUET2B~LO$^{**}$\\
× & AUET2~LO$^{**}$ AMBT1\\
× & AMBT2B~CTEQ6L1\\
× & Perugia~2010\\
× & Perugia~2011\\
\pythia~8.150 & $4C$\\
\herwigpp~2.5.1 &  UE7000-2\\
\noalign{\smallskip}\hline
\end{tabularx}
\end{table}

We would like to mention that an analytic estimate for the non-perturbative correction to the inclusive jet spectrum was proposed \cite{Dasgupta:2007wa}. It was used in \cite{Soyez:2011np} to compare the perturbative QCD calculations to the first ATLAS inclusive jet measurement \cite{Aad:2010ad}.  The values for analytic and MC-based corrections were found to be  compatible \cite{Soyez:2011np}. It would be very interesting to include the analytic evaluation of non-perturbative effects in the $\alpha_s$ determination, once the flavour decomposition of inclusive jet spectra and UE parameters are measured experimentally.

\section{\as Evaluation Procedure}
\label{Sec:EvaluationProcedure}

In this section we describe several possible procedures for determining \asZ from the inclusive jet cross section data.
First we explain how we exploit the experimental information in individual (\pt; \absy) bins, together with the propagation of the uncertainties and bin-to-bin correlations.
Then we discuss several possible averaging procedures, for combining the statistical information of the various (\pt; \absy) bins, in order to obtain an average \asZ value.
We also discuss the specific properties of these various procedures.

\subsection{Determination of \as in individual (\pt; \absy) bins}
\label{Sec:DeterminationIndBins}

An \as value is obtained in each bin of the measurement, using the theoretical prediction and the experimental cross section.
The theoretical prediction provides, in each (\pt; \absy) bin, a one-to-one mapping between \as and the inclusive jet cross section~(see previous section).
The nominal value of the measured cross section is mapped by this procedure to the nominal \as value in the corresponding bin.
All the experimental uncertainties of cross sections, together with their bin-to-bin correlations, are propagated to the determined \as values, using
pseudo-experiments~(toys), as explained in the following subsection.

\subsection{Propagation of data uncertainties and correlations}
\label{SubSec:PropUncertCorrel}

A series of pseudo-experiments, each of them including a complete set of cross section values in all the (\pt; \absy) bins, are generated using the
experimental input described in section~\ref{Sec:InputData}.

The statistical fluctuations are generated using the information provided in the published covariance matrices.
Each matrix is diagonalised, and a set of independent random numbers~(Gaussian distributions with widths given by the eigenvalues) is generated in 
the corresponding space.
The unitary matrices of eigenvectors are then used to propagate this set of random values to the physical space of the (\pt; \absy) bins.

Each nuisance parameter for a given component of the systematic uncertainties is treated as $100\%$ correlated between all the bins.
The various nuisance parameters are treated as independent with respect to each other.
The distributions of the experimental uncertainties are modelled using asymmetric Gaussian shapes, centered at zero, with $50\%$ probability of generating 
pseudo-measurements at negative or positive values.
The two widths of the asymmetric Gaussian are equal to the corresponding uncertainties~\footnote{This model of the distributions is motivated by 
the origin of the asymmetric uncertainties, due to the shape of the \pt spectra. However, as explained later, other possible models of these distributions 
have been tested.}.
In a given pseudo-experiment, the contributions of all the fluctuated nuisance parameters are added linearly to the nominal measured value.
After adding the contribution of the statistical uncertainty, the value of the pseudo-measurement is obtained.

For each pseudo-experiment, the same procedure as for the nominal measurements is applied to obtain an \as value in each bin.
This allows the propagation of all the experimental uncertainties, with their correlations and their asymmetric distributions, from the cross section level
to the \as values.

In view of the study of various possible averaging procedures, it is interesting to determine an approximate~(after the symmetrisation of uncertainties)
covariance matrix of the \as values.
The effect of the needed symmetrisation is rather small at this level, due to a partial compensation between the asymmetries of the cross section uncertainties
and the non-linear effects in the theoretical prediction~(for $\sigma(\as)$).
It is also interesting to determine a covariance matrix of the cross section measurements; however, the effect of the symmetrisation is larger at that level.

\subsection{Averaging procedures}
\label{SubSec:AveragingProcedures}

In order to study and avoid potential biases introduced by some averaging procedures, several options for performing the combination of the inputs in the various (\pt; \absy) bins have been tested
here.
All of them aim at obtaining an average value of the strong coupling constant~(\asAv).

\subsubsection{Simple average}
The first procedure consists of a simple average, where all the input \as values have the same weight~(equal to the inverse of the number of inputs).
Although the final uncertainty on this average might not be optimal, the weights of the input \as values are well behaved~(i.e. they are all in the $[0;1]$ interval, with the sum equal to unity).

\subsubsection{Weighted average}
The $2^{nd}$ procedure which has been tested consists of a weighted average, where the weights of the input \as values are proportional to the inverse of their squared total uncertainties~(obtained 
after symmetrisation).
The weights used in this procedure are also well behaved, in the sense defined above.
They also have the advantage that more precise contributions get larger weights.
A variation of this procedure has also been tested, where the weights are proportional to the inverse of the squared statistical uncertainty.
Actually, the statistical uncertainties exhibit much smaller correlations comparing to the systematic ones, motivating this additional test.

\subsubsection{Standard \chiSq minimisation}
The $3^{rd}$ averaging procedure consists of the minimisation of a ``standard'' \chiSq with correlations
\beqn
   \chiSq = (\asNomVec - \asAvVec) \cdot C^{-1} \cdot (\asNomVec - \asAvVec)^{T},
   \label{Eq:chi2TotCov}
\eeqn
with respect to \asAv.
Here, \asNomVec is the vector of nominal \as values~(obtained in individual (\pt; \absy) bins) entering the average, $C$ is their covariance matrix,
while \asAvVec is a vector containing values equal to \asAv~(to be fitted) and having the same size as \asNomVec.

It has been shown in~\cite{D'Agostini:1993uj} that this approach, although very commonly used, can yield biased average values.
Actually, in presence of (not very well understood)~strong correlations among the inputs, the average value can be even outside the range of the input values, which seems unacceptable.
Indeed, the minimisation of the function in Eq.~\ref{Eq:chi2TotCov}, with respect to \asAv~(equivalent to the minimisation of the variance of the output of a weighted average, cf. Gauss-Markov
theorem), yields~(see for example~\cite{Nakamura:2010})
\beqn
   \asAv = \frac{\bar{1} \cdot C^{-1} \cdot (\asNomVec)^{T}}{\bar{1} \cdot C^{-1} \cdot \bar{1}^{T}},
   \label{Eq:chi2minimum}
\eeqn
where $\bar{1}$ is a vector with all the entries equal to unity and the same size as \asNomVec.
Just as in the two previous methods, the average resulting from the \chiSq minimisation is a weighted mean of the input values.
However, the only constraint on these weights is that their sum equals unity.
There is no constraint on the range of the weights and, in presence of strong correlations, these weights can even be negative or larger than unity.

The explanation of the bias in this procedure, as provided in~\cite{D'Agostini:1993uj}, makes responsible the input covariance matrix because of ``the linearisation on which the usual error propagation relies''.
However, two different problems of this approach have been pointed out in~\cite{Blobel:2003wa}.
Actually, Eq.~\ref{Eq:chi2TotCov} involves the inversion of the covariance matrix, which is ill-behaved~(a small change in the covariance matrix can have an important impact on its inverse,
hence on the weights and the average).
Without special techniques, neither the inverse can be computed, nor the minimisation can be performed.
Furthermore, this \chiSq definition treats all the uncertainties as absolute~(i.e. corresponding to additive effects), whereas at least some of them~(like
the uncertainty on the luminosity etc.) should be treated as relative uncertainties~(i.e. corresponding to multiplicative effects).

\subsubsection{Modified \chiSq minimisation}
\label{SubSubSec:ModifiedChisq}
A possible solution to these problems of the ``standard'' \chiSq minimisation has been provided in~\cite{D'Agostini:1993uj}.
It consists in introducing, for each correlated systematic uncertainty, one nuisance parameter in the \chiSq definition, together with one extra constraint term.
For a normalisation uncertainty, a scaling factor is applied to both data and the uncorrelated uncertainties.
This is equivalent to introducing the corresponding nuisance parameter as a scaling factor on the theoretical prediction.
This alternative approach is used for example in~\cite{Blobel:2003wa,Pascaud:1995qs}.
For an absolute uncertainty, the nuisance parameter acts a coherent additive shift of the theoretical prediction.
However, it has been shown~(see~\cite{Botje:2001fx,ListTalk}) that, if all the systematic uncertainties correspond to~(are treated as) additive effects,
the approach using nuisance parameters is equivalent to the one of the ``standard'' \chiSq with correlations~(Eq.~\ref{Eq:chi2TotCov}).

This solution, treating the systematic uncertainties as nuisance parameters in the fit, is commonly known as ``profiling''.
It often leads to reduced systematic uncertainties, as they are constraint by the \chiSq minimisation.
However, this reduction strongly relies on the bin-to-bin correlations of the systematic uncertainties, as each nuisance parameter is treated as $100\%$ correlated between the bins.
This strong assumption~(exploited for propagating statistical information between the different phase-space regions) relies on information that is difficult to check in a particular measurement and
is therefore, in most cases, not justified.
This yields too optimistic results, as the uncertainties on the shape and correlations for each component of the systematic uncertainties should also be taken into account.
It should be pointed out that the same assumptions and uncertainty reductions are implicit in the ``standard'' \chiSq with correlations~(Eq.~\ref{Eq:chi2TotCov}).
In order to better understand this problem, one can consider an example where, in the performance studies, a given component of the systematic uncertainty has been conservatively overestimated
in some particular phase-space region.
This conservative approach would be ``exploited'' by the fitting procedure, in order to reduce the corresponding component in all the phase-space regions.

In our case, the only systematic uncertainty for which the correlations are perfectly known is the one on the normalisation~(i.e. the experimental uncertainty of the ATLAS luminosity).
This motivates the fourth combination procedure that we have tested, where the luminosity uncertainty is treated as nuisance parameter in the fit~(it is ``profiled''), while
all the other uncertainties are integrated in a partial covariance matrix.
Therefore, the \chiSq definition will include a contribution from the comparison between data and theory, involving this partial covariance matrix, plus a constraint on the luminosity shifts.

As it is straightforward to propagate a luminosity variation to the cross section, the first implementation of this combination procedure uses the \chiSq test defined as
\beqn
   && \!\!\!\!\!\!\! \chiSq_{\rm tot}\left(\sigma(\as); \beta_{\rm L}\right) = \left( \frac{\beta_{\rm L} - 1}{\epsilon_{\rm L}} \right)^{2} + \\
   && \!\!\!\!\!\!\! \left(\XsecNomVec - \XsecThVec(\asAv) \cdot \beta_{\rm L}\right) \cdot C_{{\rm -L;}\sigma}^{-1} \cdot \left(\XsecNomVec - \XsecThVec(\asAv) \cdot \beta_{\rm L}\right)^{T} \nonumber
\eeqn
Here, \XsecNomVec is the vector of nominal experimental cross section values, $\XsecThVec(\asAv)$ is the vector of theoretical cross section values computed for \asAv~(to be fitted),
$C_{{\rm -L;}\sigma}$ is the covariance matrix of the experimental cross section measurements, including all the uncertainties but the one for the luminosity,
$\beta_{\rm L}$ is the luminosity scaling factor~(ratio of the scaled to the nominal luminosity), and $\epsilon_{\rm L}$ is the luminosity relative uncertainty.
The first contribution to the total \chiSq is given by a term~($\chiSq_{\rm L}(\beta_{\rm L})$) constraining the shifts of the luminosity nuisance parameter, while 
the second~($\chiSq_{\rm d}\left(\sigma(\as); \beta_{\rm L}\right)$) comes from the comparison between data and theory at the cross section level.

As explained in subsection~\ref{SubSec:PropUncertCorrel}, the determination of the covariance matrix at the cross section level involves a larger symmetrisation, comparing to the determination
of the covariance matrix of the \as values.
This motivates a second implementation of this combination procedure, driven by the \chiSq test:
\beqn
   && \chiSq_{\rm tot}(\as; \beta_{\rm L}) = \left( \frac{\beta_{\rm L} - 1}{\epsilon_{\rm L}} \right)^{2} + \\
   && \left(\asNomVec - \asAvVec\left(\beta_{\rm L}\right) \right) \cdot C_{{\rm -L;}\as}^{-1} \cdot \left(\asNomVec - \asAvVec\left(\beta_{\rm L}\right) \right)^{T}. \nonumber
   \label{Eq:chi2asLumiProf}
\eeqn
It involves the covariance matrix of the \as values~($C_{{\rm -L;}\as}$, including all the uncertainties except the one for the luminosity), which is better defined, as explained above.
$\asAvVec\left(\beta_{\rm L}\right)$ is a vector containing values equal to $\asAv\left(\beta_{\rm L}\right)$~(the same for all the (\pt; \absy) bins) and having the same size as \asNomVec.
Here we have again the first contribution given by the term constraining the shifts of the luminosity nuisance parameter, while 
the second~($\chiSq_{\rm d}\left(\as; \beta_{\rm L}\right)$) comes from the comparison between data and theory, at the \as level.
However, as the theoretical predictions of the cross section values in the various (\pt; \absy) bins exhibit slightly different \as dependencies~(due to NLO QCD and non-perturbative effects), the
analytical dependence of \asAv on a luminosity shift can be only approximate.
In this second implementation, we use the following ansatz:
\beqn
   \asAv\left(\beta_{\rm L}\right) \approx \as \times \left( 1 + \frac{\beta_{\rm L}-1}{1.8} \right),
   \label{Eq:EffectiveAsLumiDependence}
\eeqn
which is motivated by the small size of the non-linear effects in the theoretical prediction, together with an empirical observation, as explained in section~\ref{Sec:Chi2Scan} below.

\subsubsection{Geometrical average}
The first three combination procedures described above, make use of weighted averages.
They distinguish between each other by the way how the weights of the inputs in the average are obtained.
Another~(the fifth) possibility that we have considered for performing the combination consists in a geometrical average:
\beqn
   \asAv = \left( \prod_{i=1}^{n} \asNomi \right)^{\frac{1}{n}},
\eeqn
where \asNomi runs over the n inputs of the combination.

\section{Experimental results}
\label{Sec:Results}

In this section we present the results for the \as determinations in the individual (\pt; \absy) bins, using \AKT jets with R=0.6, followed by the results of the various combination procedures.
We also discuss the issue of the disagreement between the results obtained for \AKT jets with R=0.4 and R=0.6, respectively.

\subsection{Individual (\pt; \absy) bins and their correlations}

\begin{figure*}[htbp]
\begin{center}
 \subfigure[~$45 \le \pt < 60$ GeV]{
  \includegraphics[width=8.cm]{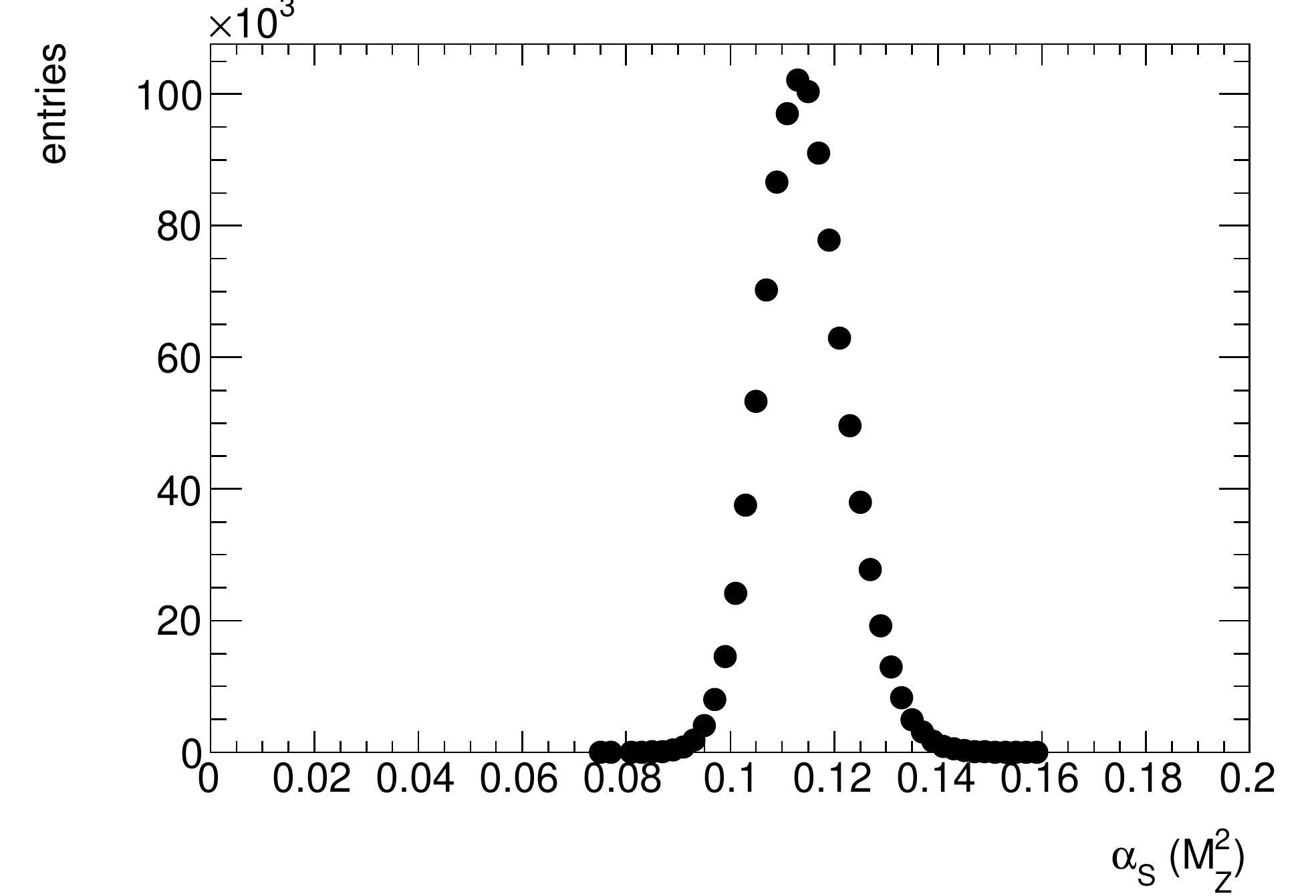}\hspace{0.4cm}
 }
 \subfigure[~$110 \le \pt < 160$ GeV]{
  \includegraphics[width=8.cm]{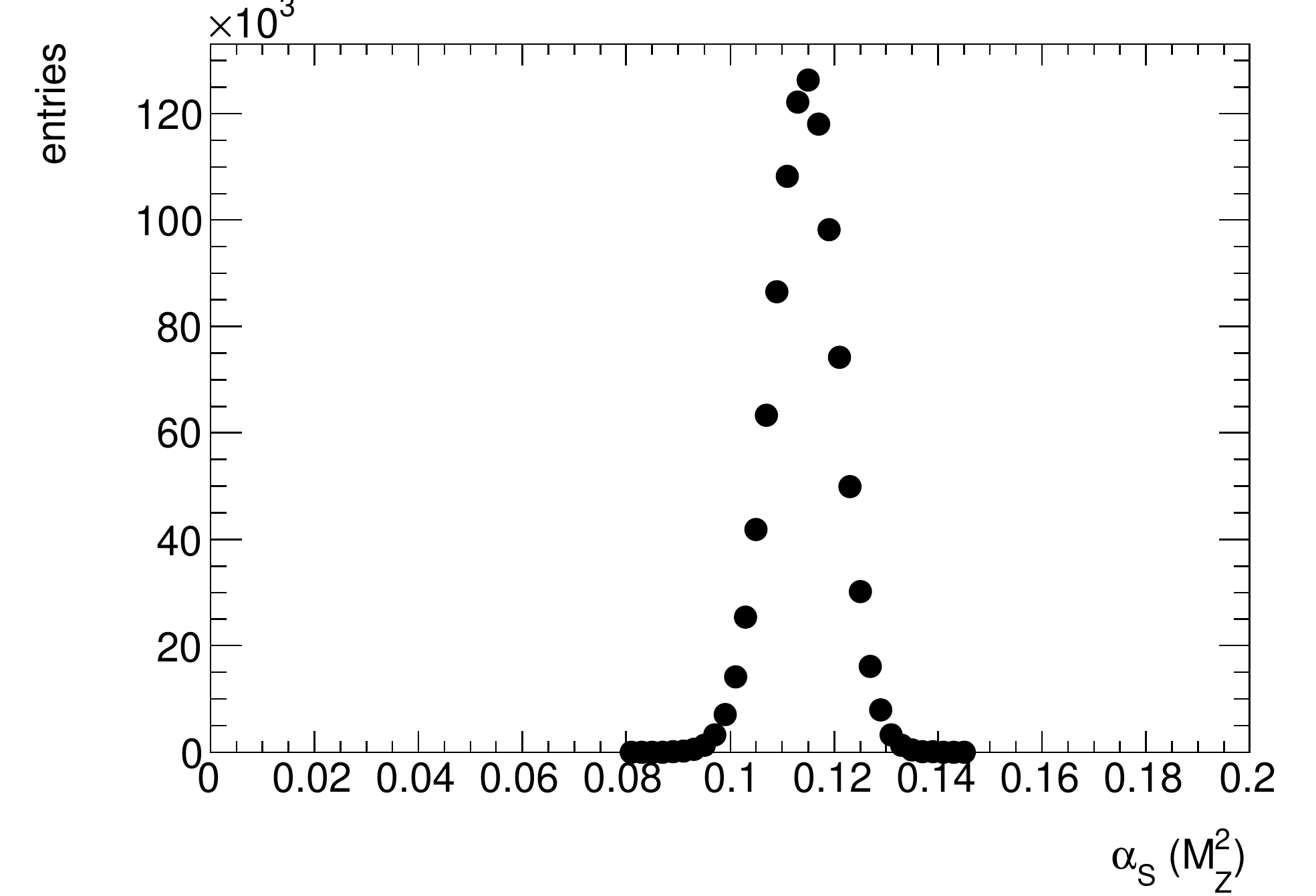}
 }\\
 \vspace{1cm}
 \subfigure[~$310 \le \pt < 400$ GeV]{
  \includegraphics[width=8.cm]{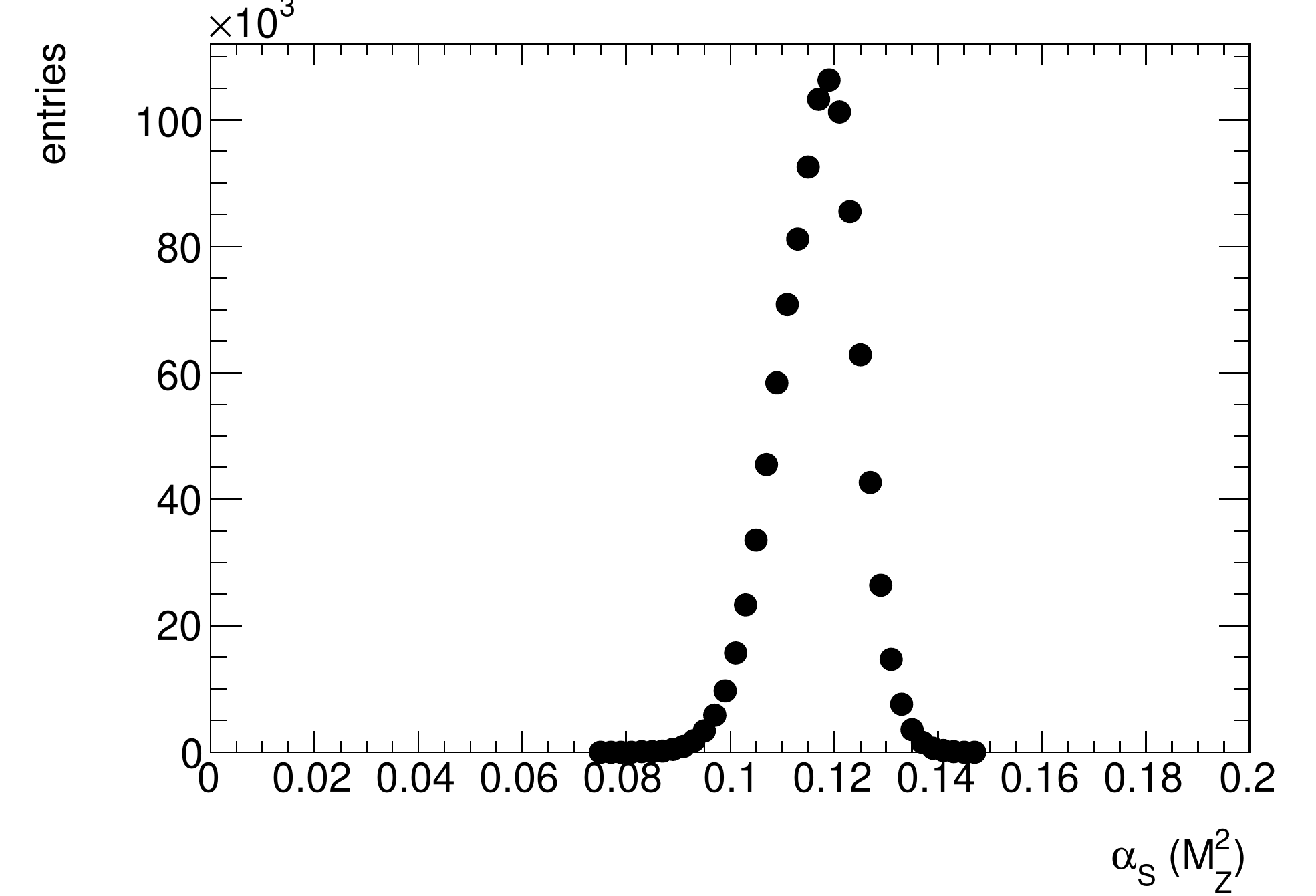}\hspace{0.4cm}
 }
 \subfigure[~$500 \le \pt < 600$ GeV]{
  \includegraphics[width=8.cm]{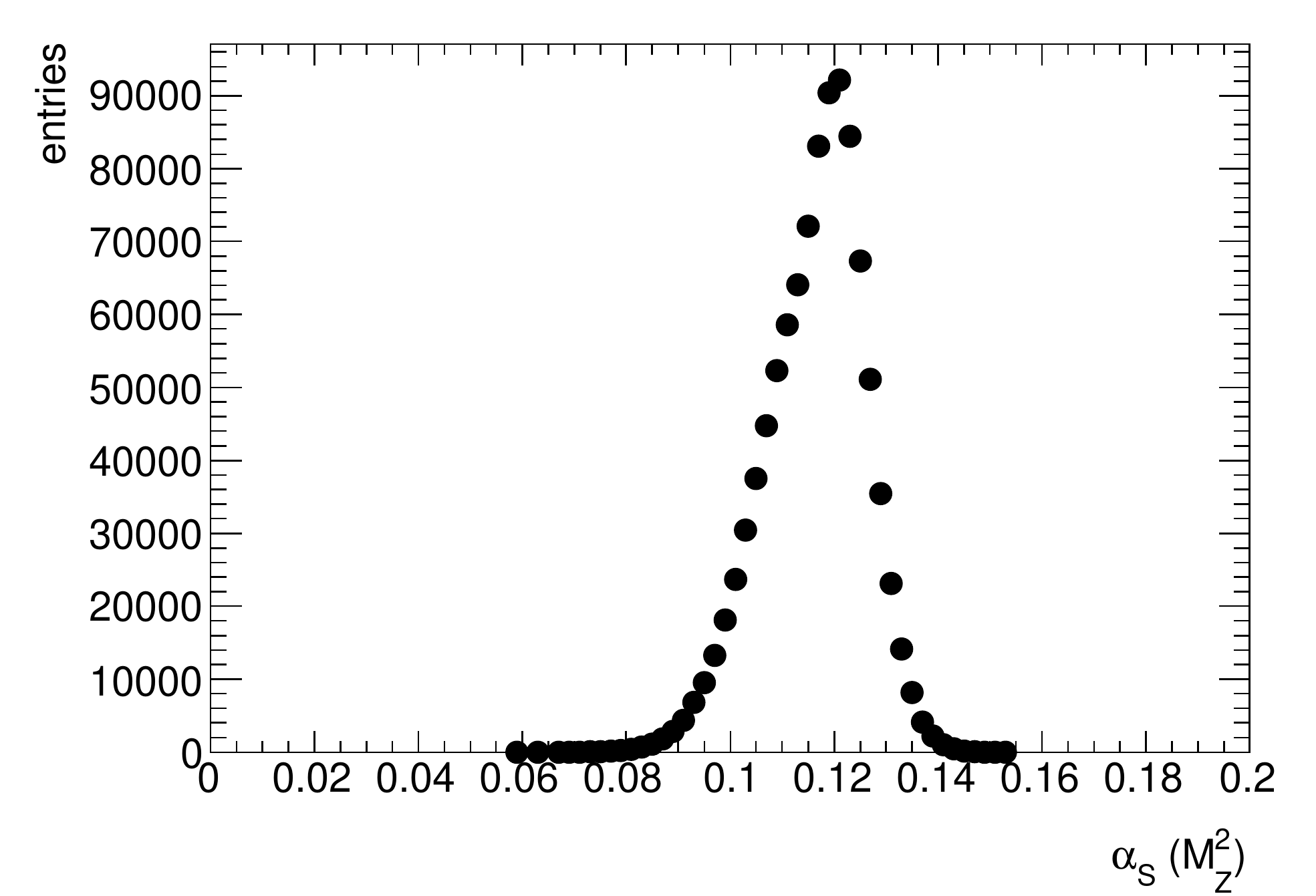}
 }
\end{center}
\vspace{-0.5cm}
\caption[.]{ 
  Examples of distributions of \asZ, as obtained from $10^6$ pseudo-experiments, for \AKT jets with R=0.6, in the central rapidity region~($0 \le \absy < 0.3$),
  in several \pt bins.
}
\label{Fig:asDistribPtEtaBin}
\end{figure*}

Following the prescription described in section~\ref{Sec:DeterminationIndBins}, a nominal \as value has been determined in each (\pt; \absy) bin, while the statistical and systematic
uncertainties, together with their correlations, have been propagated using pseudo-experiments.
Fig.~\ref{Fig:asDistribPtEtaBin} shows a few examples of \asZ distributions, as obtained from $10^6$ pseudo-experiments, in the central \absy region, in four different \pt bins.
Some asymmetries can be seen in these distributions.
They are due to the asymmetries in the data uncertainties and to the non-linear dependence of the theoretical cross sections on \as~(from the theoretical calculation and its extrapolation to
low and large \as values, outside the region where the \as scan is available).

For very low and very large \pt values, the total data uncertainties are relatively large.
This is mainly due to the JES uncertainty, while the statistical uncertainties are also large at high \pt.
In these regions of the \pt spectra, a fluctuation of the cross section by $\pm 1\sigma$~(or more) yields \as values for which the theoretical prediction is not reliable.
Actually, the PDF determinations are not reliable for the corresponding large or low \as values, while the perturbative calculations also exhibit problems at large \as.
It is for this reason that we discard the \pt bins with large experimental uncertainties in the final \as average computation, and we estimate a systematic uncertainty due to this choice.
Because of their large uncertainties, these bins would in any case have a small impact on the average, therefore the estimated uncertainty associated to dropping these bins is small.
A number of 42 (\pt; \absy) bins are used in the computation of the various averages.
The corresponding ($[\ptmin;\ptmax]$; $[\absy_{\rm min};\absy_{\rm max}]$) ranges, as well as the labels/numbers used for these bins, are given in table~\ref{Tab:pTyBins}.
\begin{table}[t]
 \caption[.]{ \label{Tab:pTyBins}
   \pt ranges~($[\ptmin;\ptmax]$) used for the \as determination in various \absy intervals~($[\absy_{\rm min};\absy_{\rm max}]$).
   The third column indicates the label used for the \absy bins, while the last column gives the range of numbers assigned to the corresponding 
   (\pt; \absy) bins~(ordered in increasing \pt and \absy).
}
\setlength{\tabcolsep}{0.0pc}
\begin{tabularx}{\columnwidth}{@{\extracolsep{\fill}}cccc}
\hline\noalign{\smallskip}
 $[\absy_{\rm min};\absy_{\rm max}]$  &  $[\ptmin;\ptmax] (\rm GeV)$  &  \absy bin label  &  bin numbers \\
\noalign{\smallskip}\hline\noalign{\smallskip}
 $[0;0.3]$    & [45;600]  & $\absy_{1}$ &  1-10 \\
 $[0.3;0.8]$  & [45;600]  & $\absy_{2}$ & 11-20 \\
 $[0.8;1.2]$  & [45;500]  & $\absy_{3}$ & 21-29 \\
 $[1.2;2.1]$  & [45;310]  & $\absy_{4}$ & 30-36 \\
 $[2.1;2.8]$  & [60;210]  & $\absy_{5}$ & 37-40 \\
 $[2.8;3.6]$  & [80;110]  & $\absy_{6}$ &   41  \\
 $[3.6;4.4]$  & [80;110]  & $\absy_{7}$ &   42  \\
\noalign{\smallskip}\hline
\end{tabularx}
\end{table}

\begin{figure*}[htbp]
\begin{center}
 \subfigure[~Covariance matrix of cross section uncertainties.]{
   \includegraphics[width=8.cm]{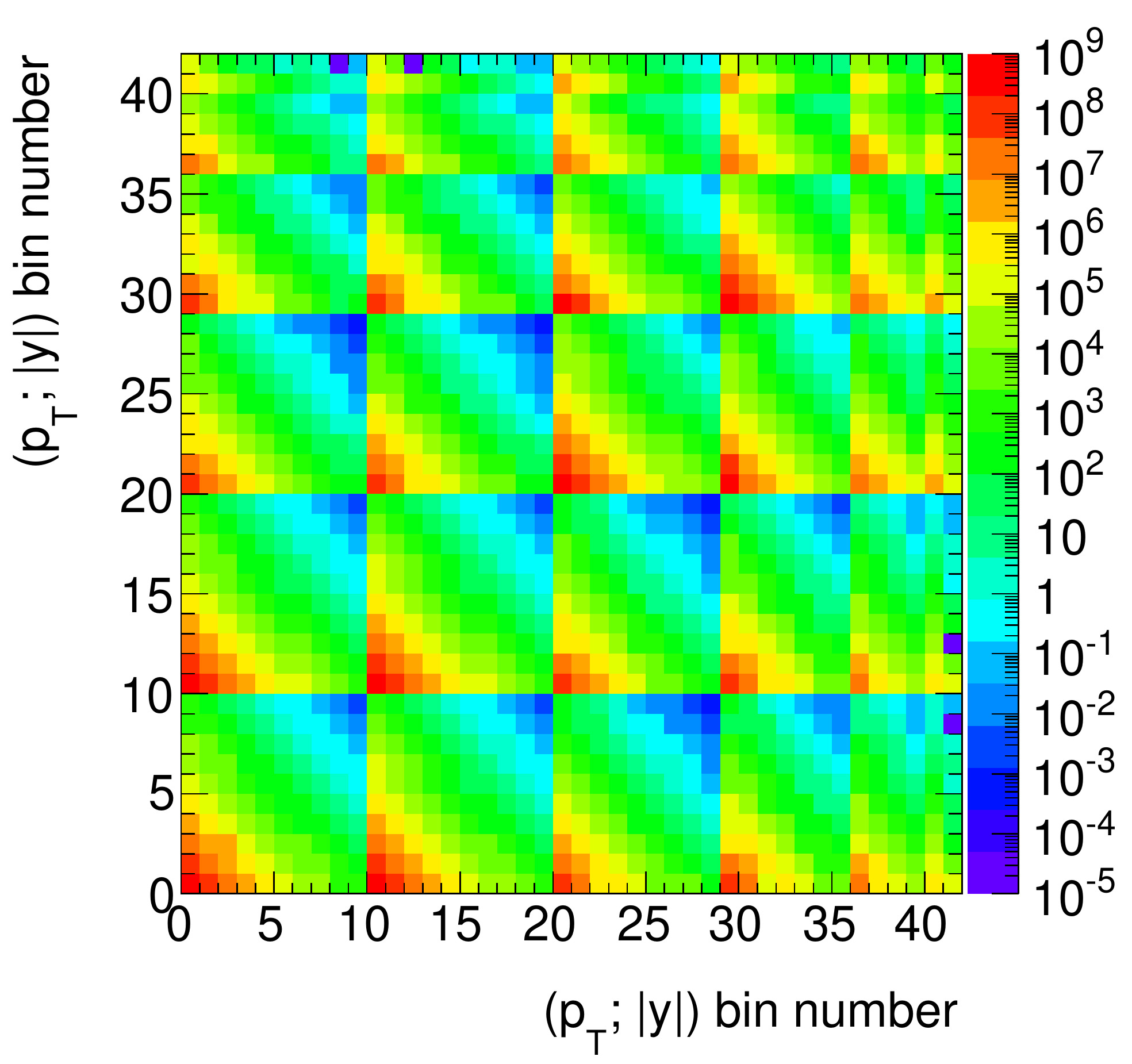}\hspace{0.4cm}
   \label{Fig:XsecCovMatrices}
 }
 \subfigure[~Correlation matrix of cross section uncertainties.]{
   \includegraphics[width=8.cm]{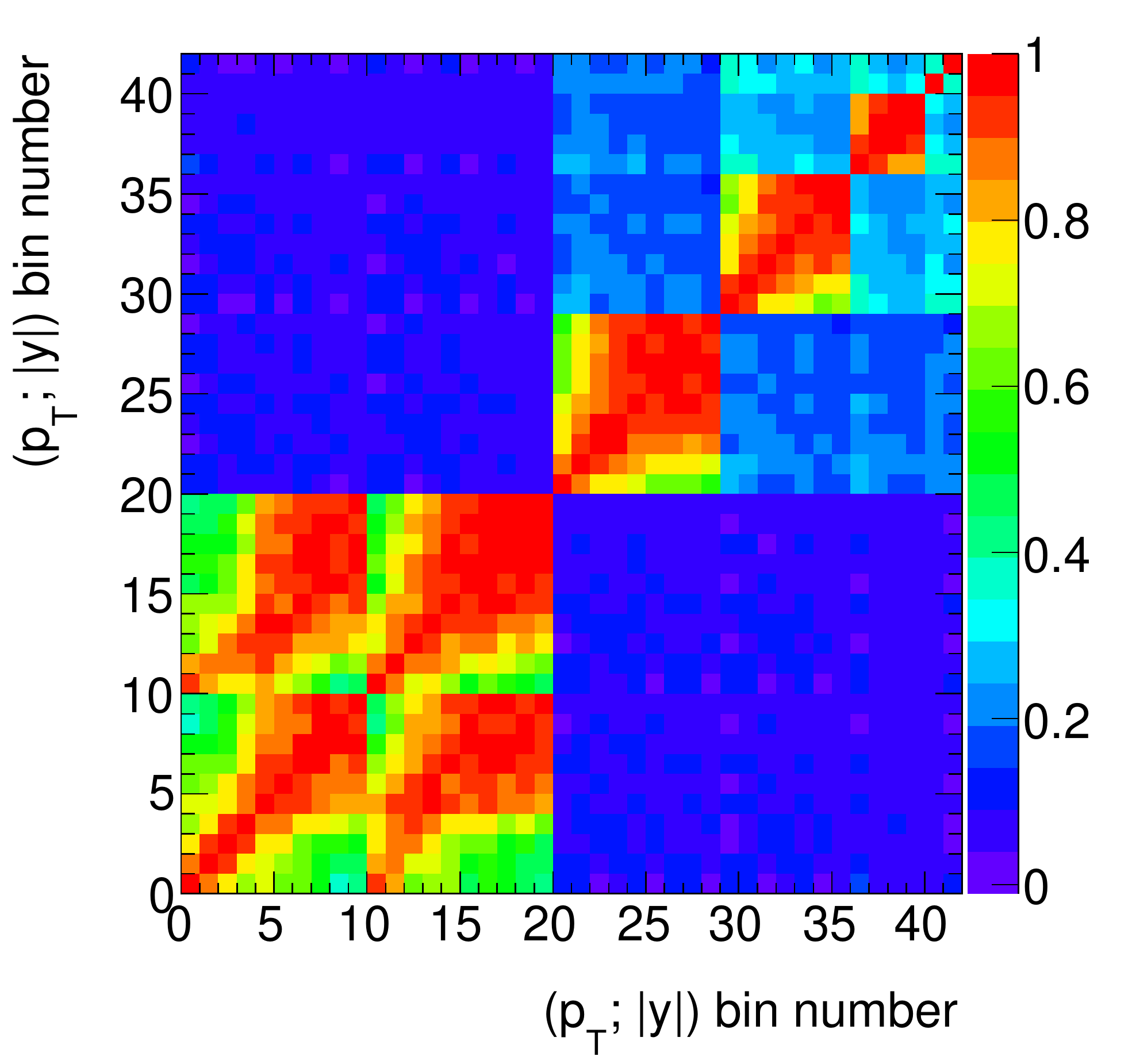}
   \label{Fig:XsecCorrMatrices}
 } \\
\vspace{1.cm}
 \subfigure[~Covariance matrix of \as uncertainties.]{
   \includegraphics[width=8.cm]{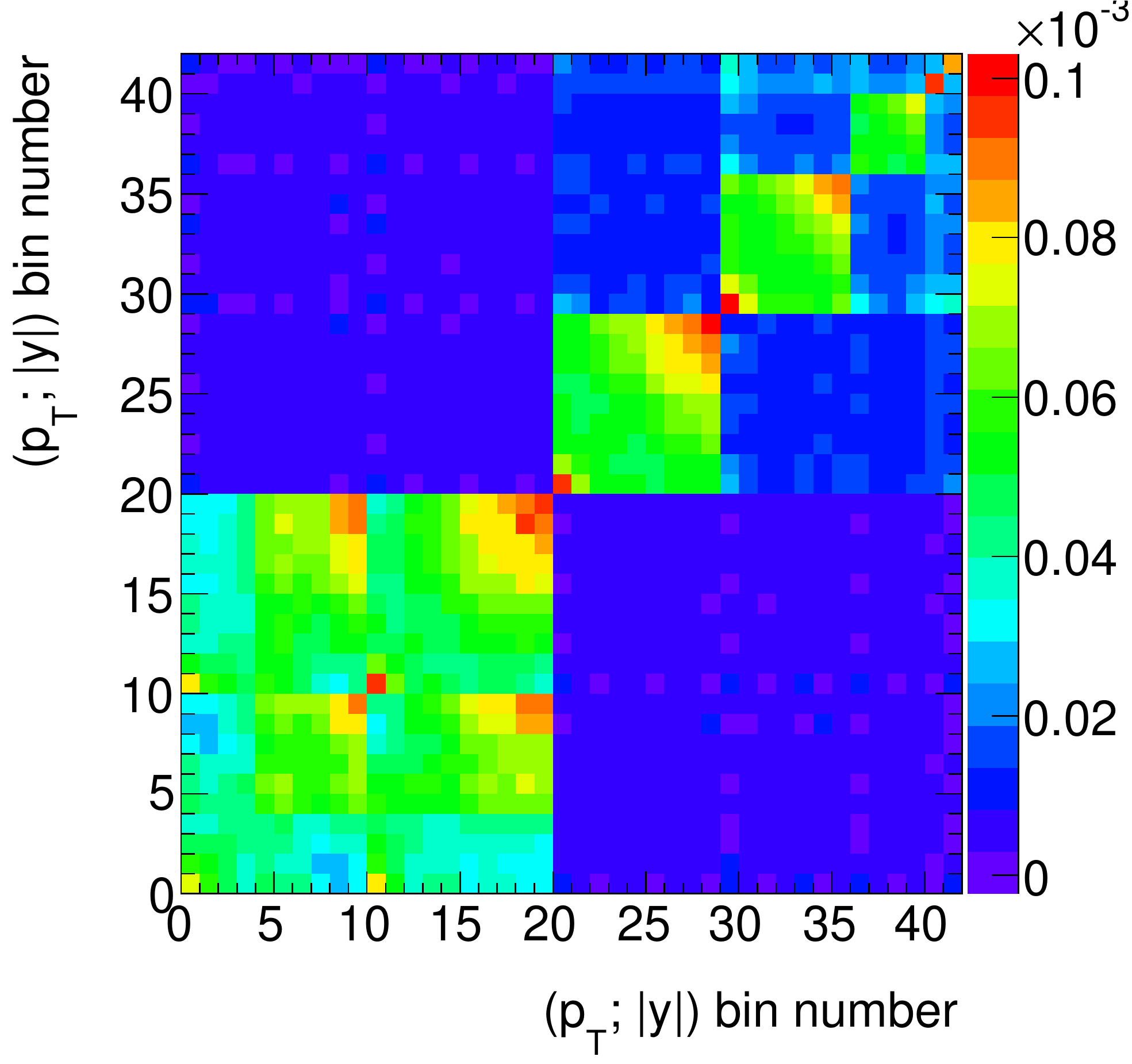}\hspace{0.4cm}
   \label{Fig:asCovMatrices}
 }
 \subfigure[~Correlation matrix of \as uncertainties.]{
   \includegraphics[width=8.cm]{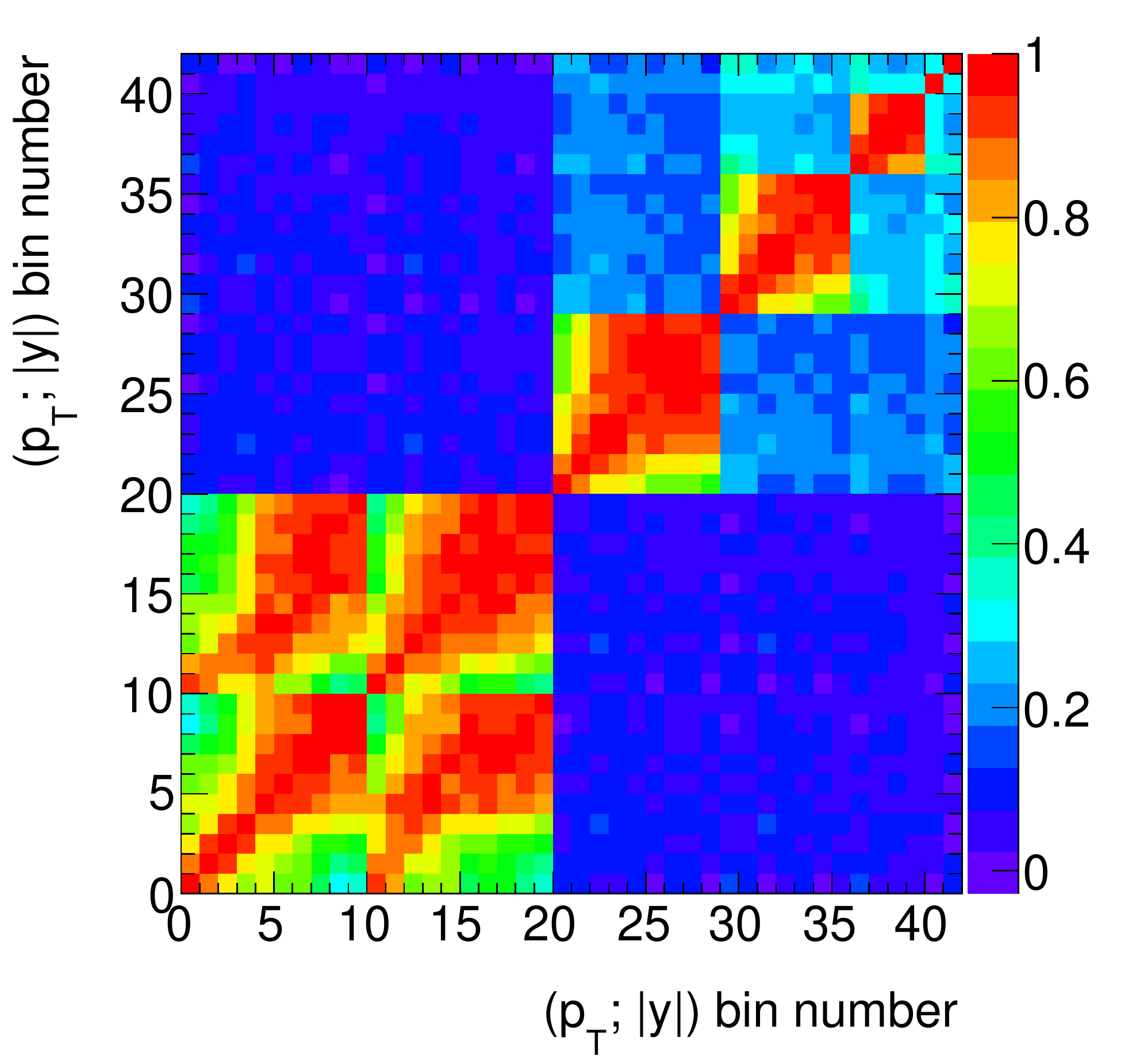}
   \label{Fig:asCorrMatrices}
 }
\end{center}
\vspace{-0.5cm}
\caption[.]{ 
  Covariance~(left) and correlation~(right) matrices of the cross section uncertainties~(top) and \as uncertainties~(bottom), in the (\pt; \absy) bins used for the final \as average determination.
  The (\pt; \absy) bins are numbered as indicated in table~\ref{Tab:pTyBins}.
}
\label{Fig:asXsecCovCorrMatrices}
\end{figure*}

The same sets of pseudo-experiments have been used to determine the (total) covariance and correlation matrices, for the cross sections and \as 
values respectively~(see Fig.~\ref{Fig:asXsecCovCorrMatrices}), in the (\pt; \absy) bins used for the final combination.
Although, as explained above, a slightly larger symmetrisation is needed in order to compute these matrices at the cross section level, it is interesting to remark that the correlation matrices
at the \as and cross section level are very similar.

\subsection{Results with a simple average}

The combination using a simple average~(SA) of all the input \as values yields
\beqns
   \as^{\rm SA} = 0.1149^{+0.0047}_{-0.0048},
\eeqns
where the quoted uncertainty is experimental only.
These uncertainties are obtained by computing~(from the pseudo-experiments) the variances of the positive and, respectively, negative fluctuations with respect to the nominal value.
These uncertainties are rather symmetric, although the asymmetries are slightly larger when computing the values in individual (\pt; \absy) bins, or even the simple averages of the \pt bins in
some particular \absy bin.

\subsection{Results with a weighted average}
\label{Sec:ResultsWeightedAverage}

While in the previous combination method the same weight was assigned to all the input \as values, here we benefit more from precise measurements, using weights proportional to
the inverse of the squared total uncertainty.
However, the gain in precision is not very big, because the various measurements that are used have rather similar precisions.
Actually, all the corresponding weights are in the range between 0 and 1, the ratio of the largest to the smallest weight being $\approx 2.5$.

\begin{figure*}[htbp]
\begin{center}
  \subfigure[~$0 \le \absy < 0.3$]{
    \includegraphics[width=7.5cm]{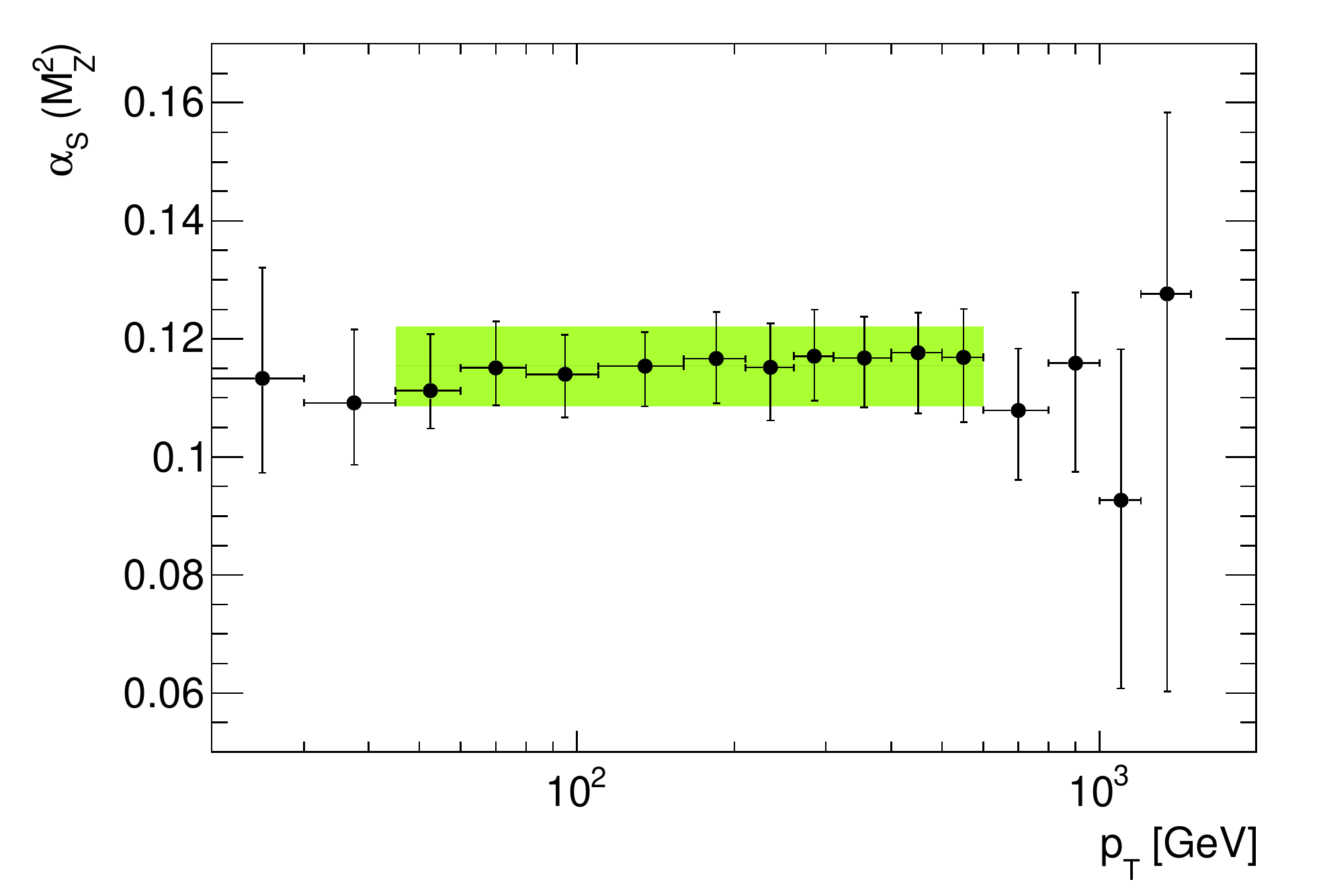}\hspace{0.4cm}
  }
  \subfigure[~$0.3 \le \absy < 0.8$]{
    \includegraphics[width=7.5cm]{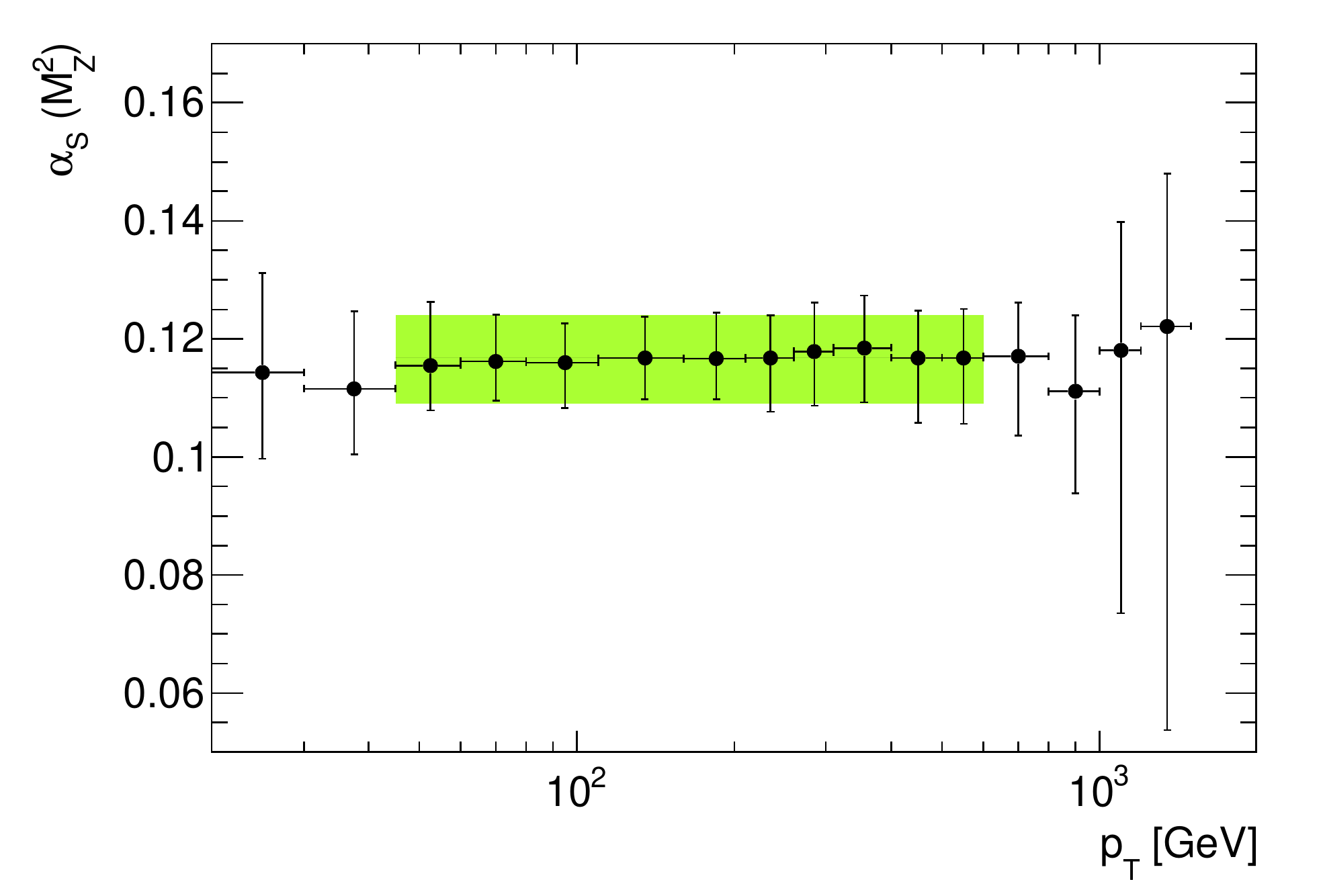}
  }\\
\vspace{0.cm}
  \subfigure[~$0.8 \le \absy < 1.2$]{
    \includegraphics[width=7.5cm]{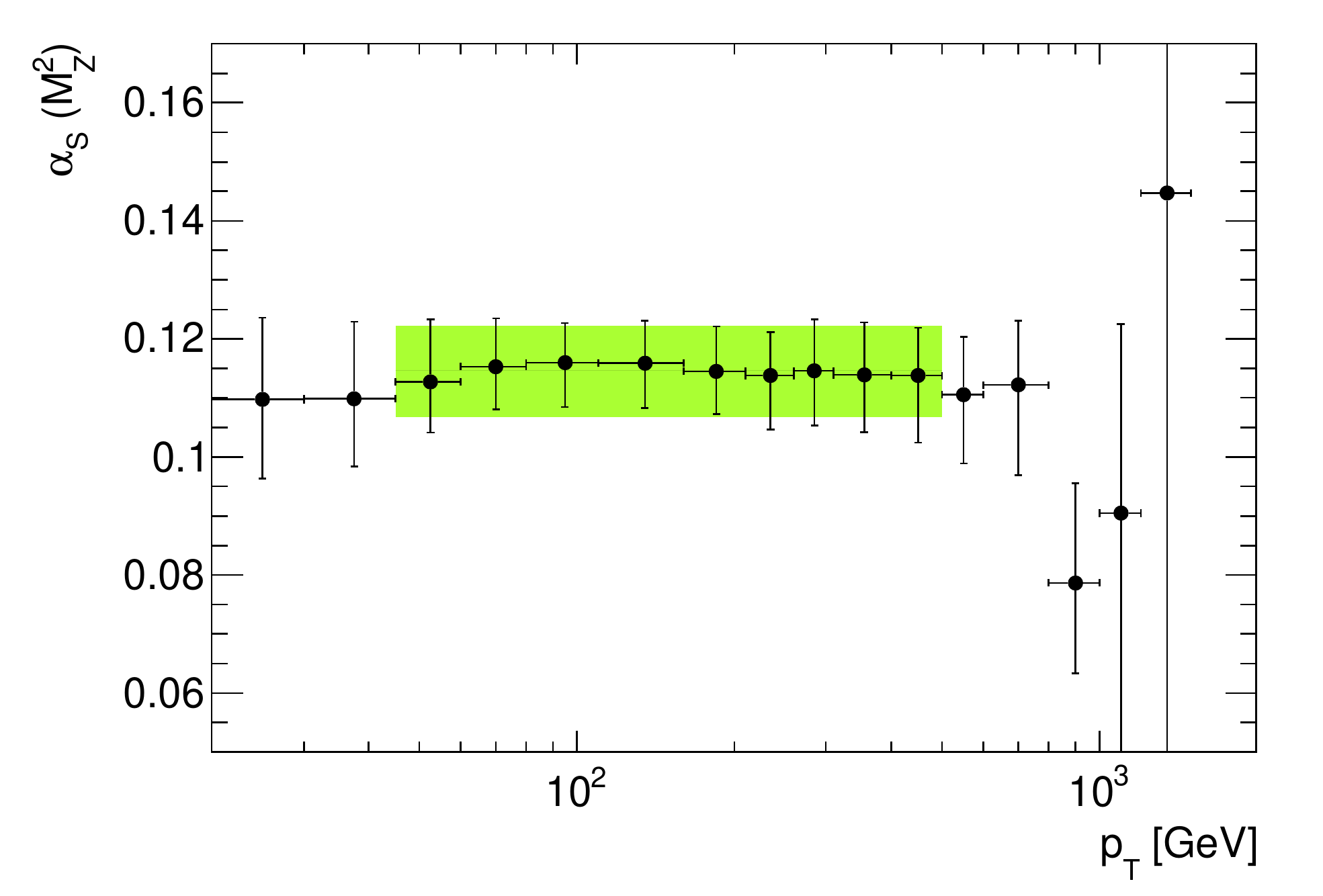}\hspace{0.4cm}
  }
  \subfigure[~$1.2 \le \absy < 2.1$]{
    \includegraphics[width=7.5cm]{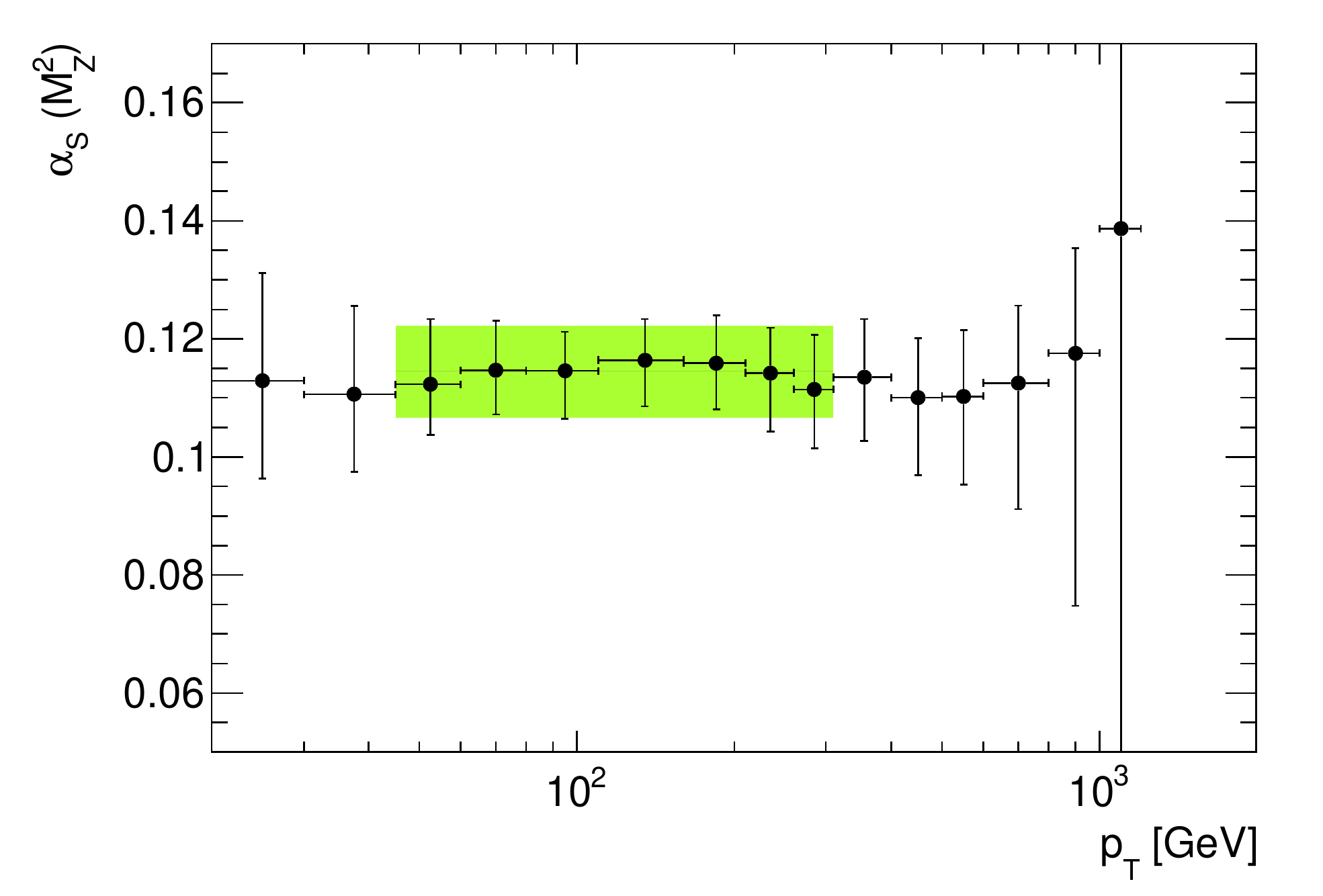}
  }\\
\vspace{0.cm}
  \subfigure[~$2.1 \le \absy < 2.8$]{
    \includegraphics[width=7.5cm]{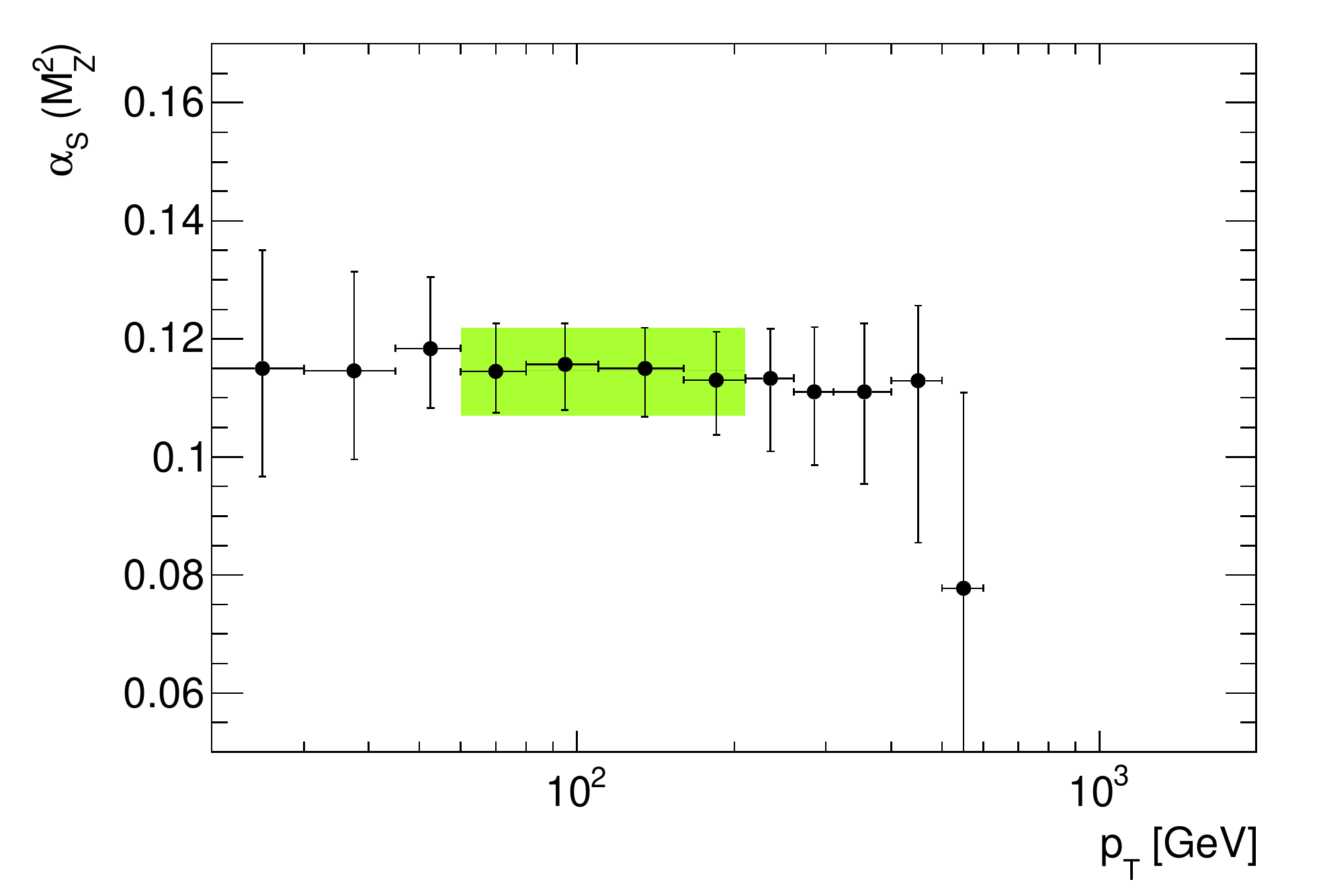}\hspace{0.4cm}
  }
  \subfigure[~$2.8 \le \absy < 3.6$]{
    \includegraphics[width=7.5cm]{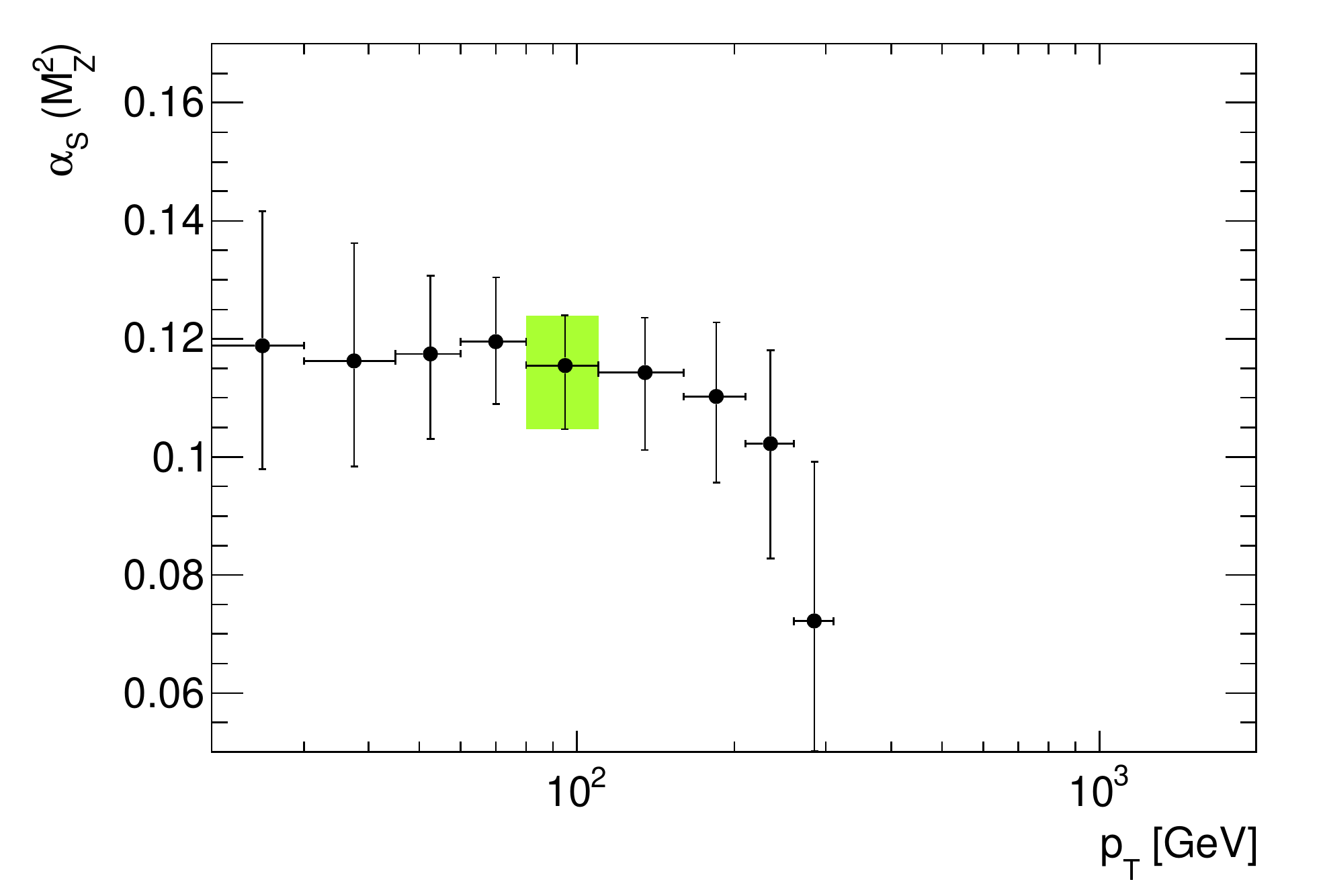}
  }\\
\vspace{0.cm}
  \subfigure[~$3.6 \le \absy < 4.4$]{
    \includegraphics[width=7.5cm]{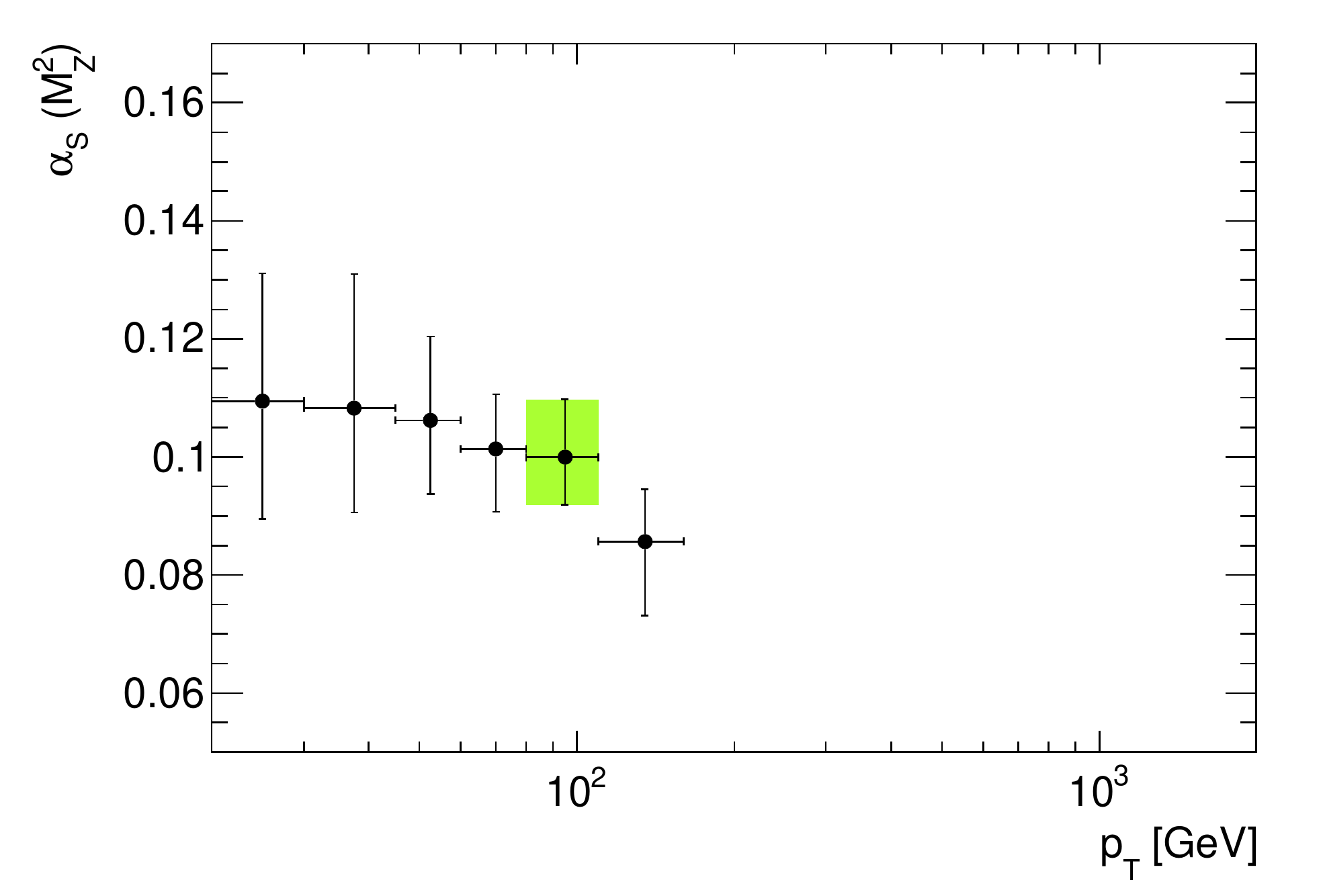}
  }
\end{center}
\vspace{-0.5cm}
\caption[.]{ 
  \asZ values~(points, with asymmetric uncertainties) obtained from the ATLAS inclusive jet cross section, in all the available (\pt; \absy) bins.
  The different figures correspond to increasing \absy values, from left to right and from top to bottom.
  The green band indicates the weighted averages obtained in the corresponding \absy bins, and it covers only the bins which are actually used for its determination~(see text).
}
\label{Fig:asWeightedAvEtaBins}
\end{figure*}

Computing this weighted average~(WA) in the individual \absy intervals given in table~\ref{Tab:pTyBins}, we obtain
\beqns
   \as^{\rm WA}{\left(\absy_1\right)} =  0.1155^{+0.0067}_{-0.0069}, \\
   \as^{\rm WA}{\left(\absy_2\right)} =  0.1167^{+0.0073}_{-0.0077}, \\
   \as^{\rm WA}{\left(\absy_3\right)} =  0.1147^{+0.0075}_{-0.0079}, \\
   \as^{\rm WA}{\left(\absy_4\right)} =  0.1145^{+0.0076}_{-0.0080}, \\
   \as^{\rm WA}{\left(\absy_5\right)} =  0.1147^{+0.0072}_{-0.0077}, \\
   \as^{\rm WA}{\left(\absy_6\right)} =  0.1155^{+0.0085}_{-0.0108}, \\
   \as^{\rm WA}{\left(\absy_7\right)} =  0.1000^{+0.0098}_{-0.0080}.
\eeqns
Fig.~\ref{Fig:asWeightedAvEtaBins} shows the \asZ values obtained in the different (\pt; \absy) bins, as well as the averages obtained in the corresponding \absy bins.
The precision of the average in each \absy bin is similar to the precision of the values obtained in the corresponding~(input) \pt bins.
This is due to the strong correlations between~(and the similar size of) the uncertainties in the \pt bins used for the average inside a given \absy bin.

Performing the combination of all the 42 (\pt; \absy) bins, one obtains
\beqn
   \as^{\rm WA} = 0.1151^{+0.0047}_{-0.0047}.
\label{Eq:NominalWeightedAverage}
\eeqn
The experimental uncertainty of this average is about $30\%$ smaller compared to the one obtained in the most precise individual \absy bin~(i.e. the central bin).
This gain in precision is due to the rather small correlations of the systematic uncertainties, between the central and the forward region~(see Fig.~\ref{Fig:asXsecCovCorrMatrices}).
\begin{figure}[htbp]
\begin{center}
\includegraphics[width=8.7cm]{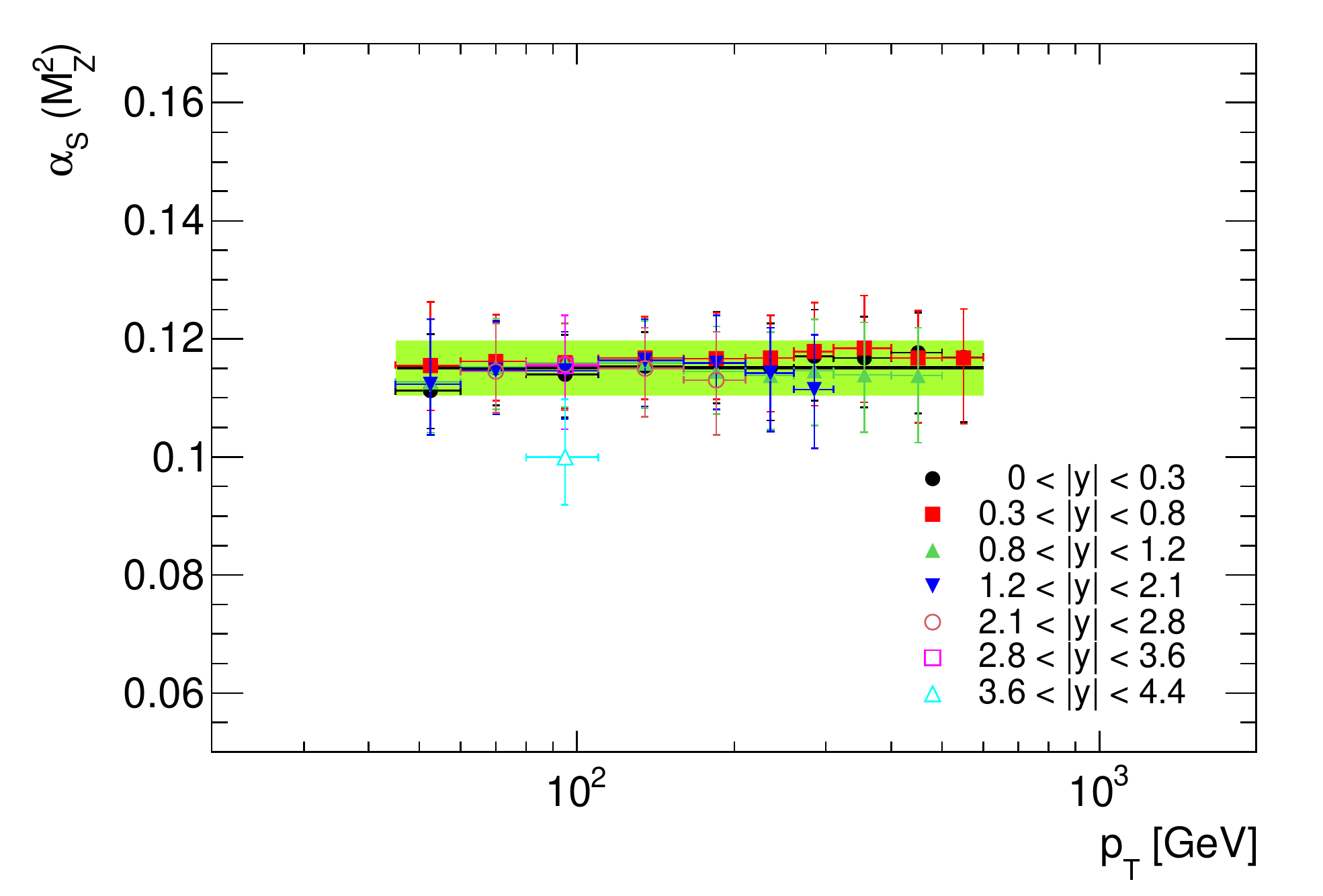}
\end{center}
\vspace{-0.5cm}
\caption[.]{ 
  Global weighted \asZ average~(green band) together with the \asZ values~(with asymmetric uncertainties) obtained in all the (\pt; \absy) bins used in the combination.
}
\label{Fig:asWeightedAvAllEta}
\end{figure}
Fig.~\ref{Fig:asWeightedAvAllEta} shows the global \asZ average compared to the values obtained in individual (\pt; \absy) bins and used in the combination.

The result of this average, together with the covariance matrix of the uncertainties for the \as values~(Fig.~\ref{Fig:asCovMatrices}),
have been inserted into Eq.~\ref{Eq:chi2TotCov}, yielding a $\chi^{2}/{\rm dof}=0.54$~(for $41$ degrees of freedom).
Using the same \asAv inserted in the theoretical prediction of the cross section, together with the covariance matrix of the uncertainties for the cross section
values~(Fig.~\ref{Fig:XsecCovMatrices}), yields, by comparison with the experimental cross section, a $\chi^{2}/{\rm dof}=0.52$.
In the individual \absy bins, where more than one \pt bin is used in the average, the $\chi^{2}/{\rm dof}$ obtained using the corresponding weighted averages and covariance matrices
are between $0.21$ and $1.39$.
Taking into account the bin-to-bin correlations is very important for the determination of these $\chi^2$ values.
Ignoring the correlations would actually yield $\chi^2$ values smaller by more than one order of magnitude.

An alternative averaging procedure has been tested, where the weights are obtained using only the statistical uncertainties~(which exhibit rather small correlations),
together with the uncorrelated components of the systematic uncertainty.
However, the full information on all the uncertainties and correlations of the ATLAS measurement is used in pseudo-experiments, for the error propagation.
The result obtained through this procedure is very similar to the previous ones~(i.e. $0.1154 ^{+0.0047}_{-0.0048}$).

The results~(central values and uncertainties) obtained using the simple average and these variations of the weighted average are consistent,
as expected for a set of inputs with similar uncertainties.

We have also tested the stability of these results when using non-Gaussian distributions for the experimental systematic uncertainties.
When using log-normal distributions, the nominal values obtained from the averages in individual \absy bins are very similar to the ones using Gaussian distributions,
while some small changes are seen in the asymmetric uncertainties.
For the result of the combination in all the (\pt; \absy) bins, both the nominal value and the uncertainties are very similar to the ones obtained for Gaussian distributions.

\subsection{Minimisation of a ``standard'' $\chi^{2}$ with correlations}

This procedure minimises a $\chi^2$ with correlation~(see Eq.~\ref{Eq:chi2TotCov}), using the covariance matrix in Fig.~\ref{Fig:asCovMatrices} as input, and yields
\beqns
   \as^{ \chi^2_{\rm min}} = 0.1165^{+0.0033}_{-0.0033}.
\eeqns
This result has a smaller experimental uncertainty compared to the previous averaging procedure.
This is not surprising, since the Gauss-Markov theorem guarantees it to have the smallest variance, among all the weighted averages with the sum of weights equal to unity~(see
section~\ref{SubSec:AveragingProcedures}).
However, the central value obtained here is questionable, since the weights of this average strongly rely on the exact knowledge of the bin-to-bin correlations and are not well behaved.
Indeed, a large fraction~(almost half) of the weights of the individual \pt bins in this average are smaller than zero.
The $\chi^{2}/{\rm dof}$ in this case is about $0.53$, slightly smaller comparing to the one in the previous weighted average~(as expected).

Looking at the similar averages computed in individual \absy bins, they are in the range between $0.1100$ and $0.1243$, several of them being outside the range of the corresponding input values.
This is due to the same bad behaviour of the weights, some of which are negative, while some others are larger than one.
At the same time, the $\chi^{2}/{\rm dof}$ computed in the individual \absy bins~(where more than one \pt bin is used in the average) are between $0.20$ and $0.93$.

\subsection{Simultaneous fit of \asZ and the luminosity, using a modified \chiSq}
\label{Sec:Chi2Scan}

As explained in section~\ref{SubSubSec:ModifiedChisq}, two modified \chiSq definitions~(at the cross section and \as level, respectively) were introduced,
treating the luminosity uncertainty as nuisance parameter.
The main goal here is to test somewhat better behaved definitions of the \chiSq with correlations.

An approximate effective dependence of \asAv on luminosity shifts is needed in order to define the $\chi^{2}$ at the \as level, given in Eq.~\ref{Eq:chi2asLumiProf}.
In order to establish this dependence, a global relative shift corresponding to one standard uncertainty of the luminosity was performed, for all the cross section measurements.
The first weighted average in section~\ref{Sec:ResultsWeightedAverage}, using the total experimental uncertainty in each (\pt; \absy) bin, was rederived.
The induced variation of the weighted average yielded, within a linear approximation, the effective $\asAv\left(\beta_{\rm L}\right)$ dependence~(see Eq.~\ref{Eq:EffectiveAsLumiDependence}).

\begin{figure*}[htbp]
\begin{center}
\includegraphics[width=8.cm]{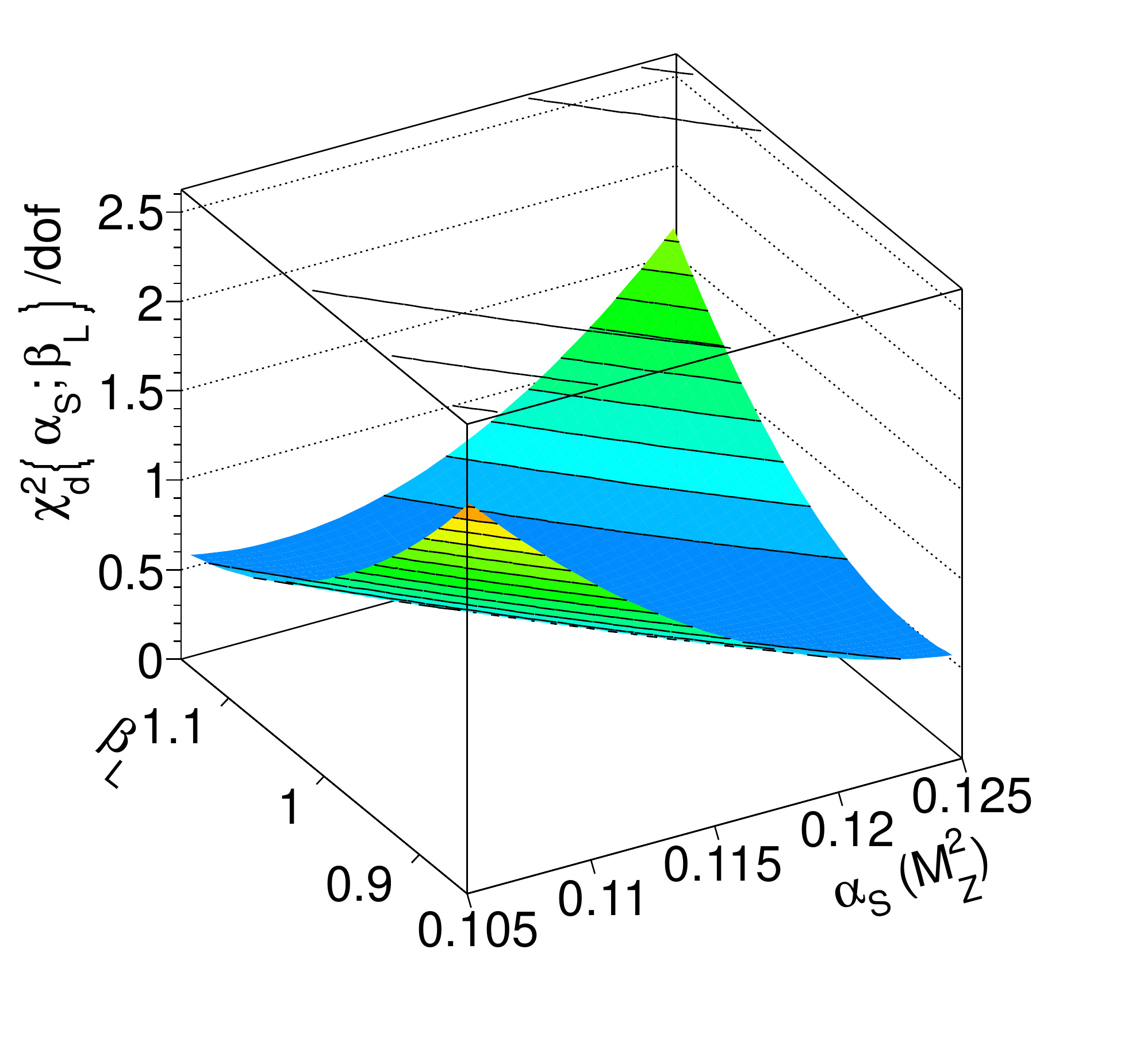}\hspace{0.4cm}
\includegraphics[width=8.cm]{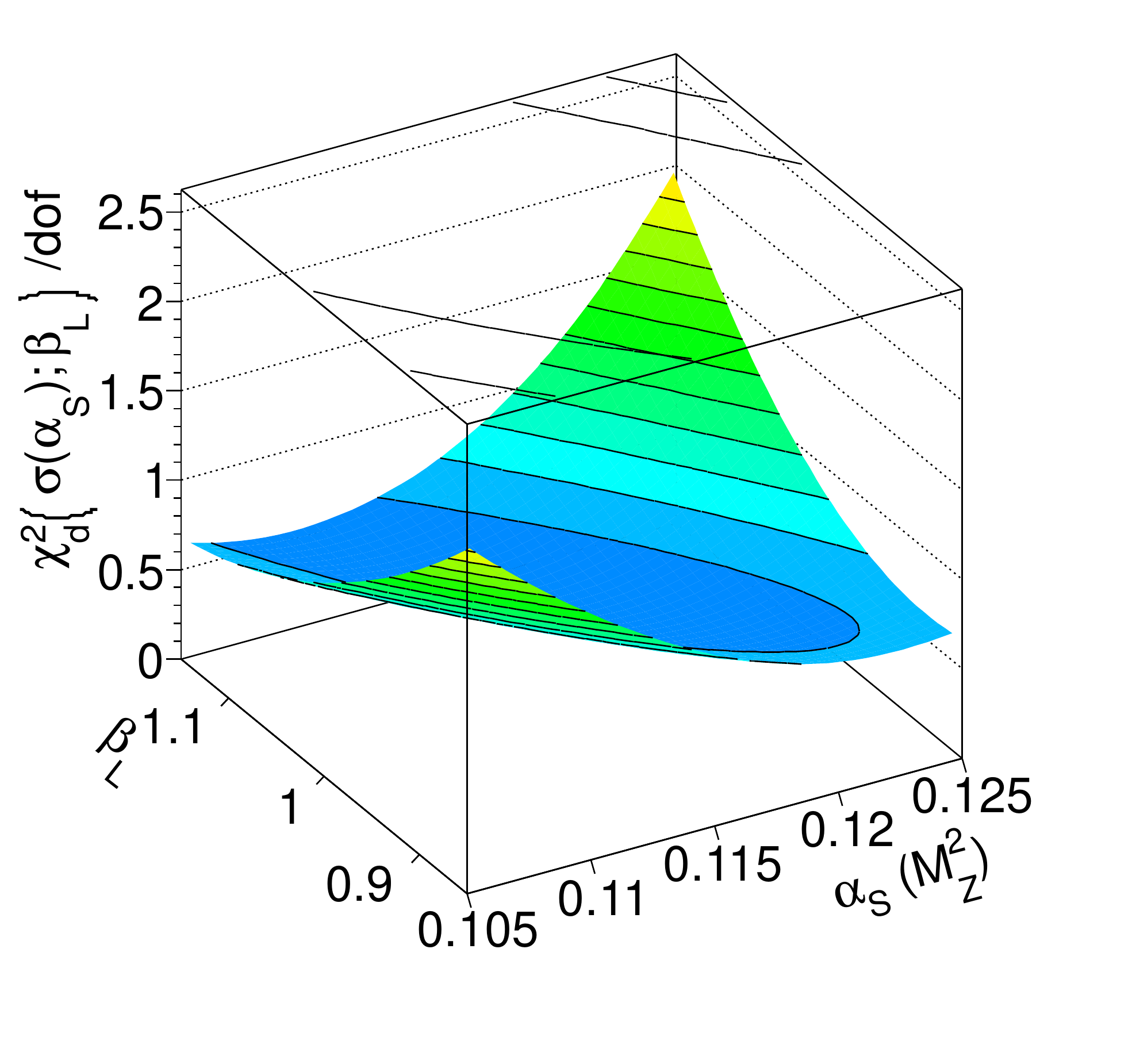}\\
\vspace{0.1cm}
\includegraphics[width=8.cm]{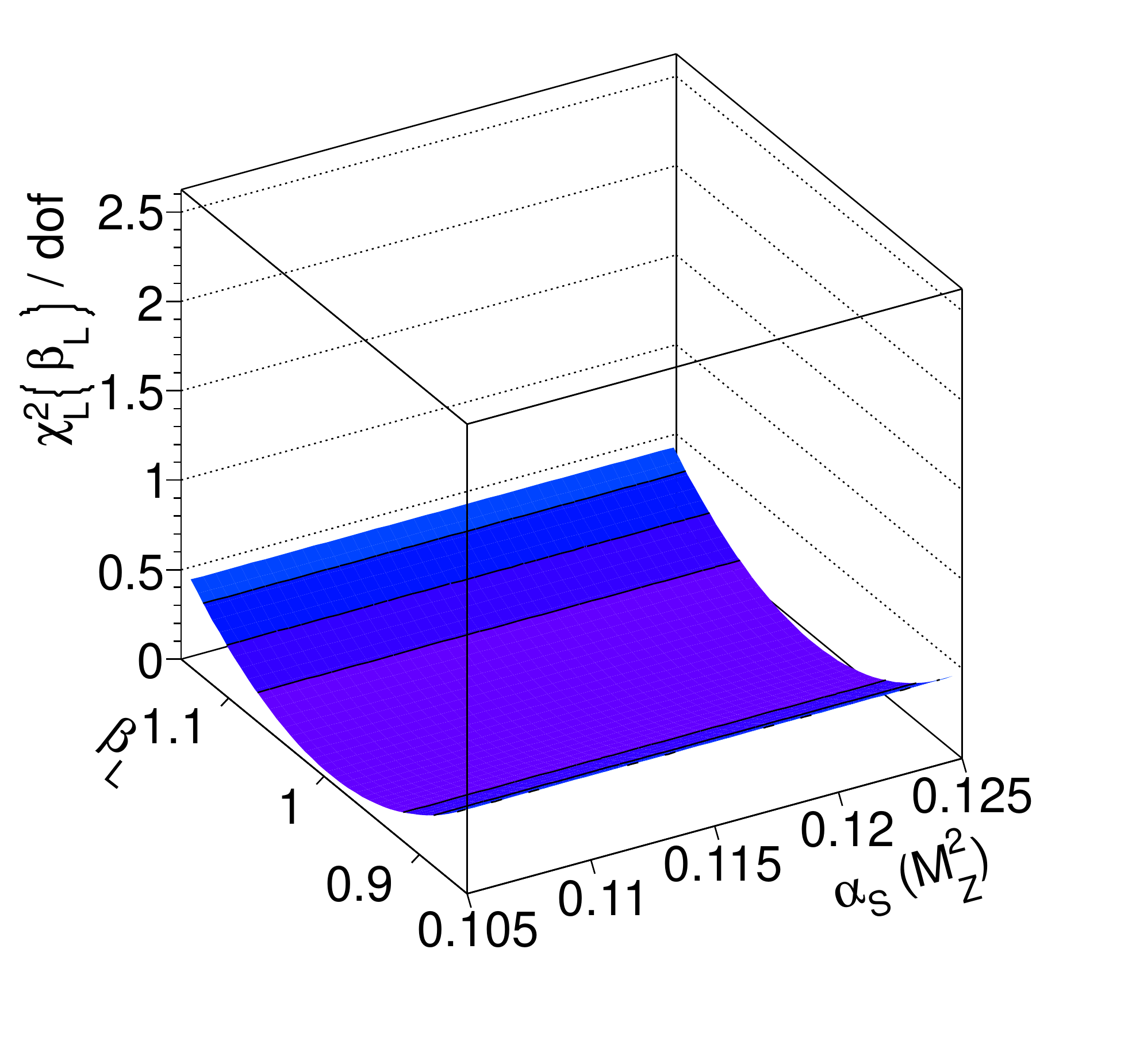}\\
\vspace{0.1cm}
\includegraphics[width=8.cm]{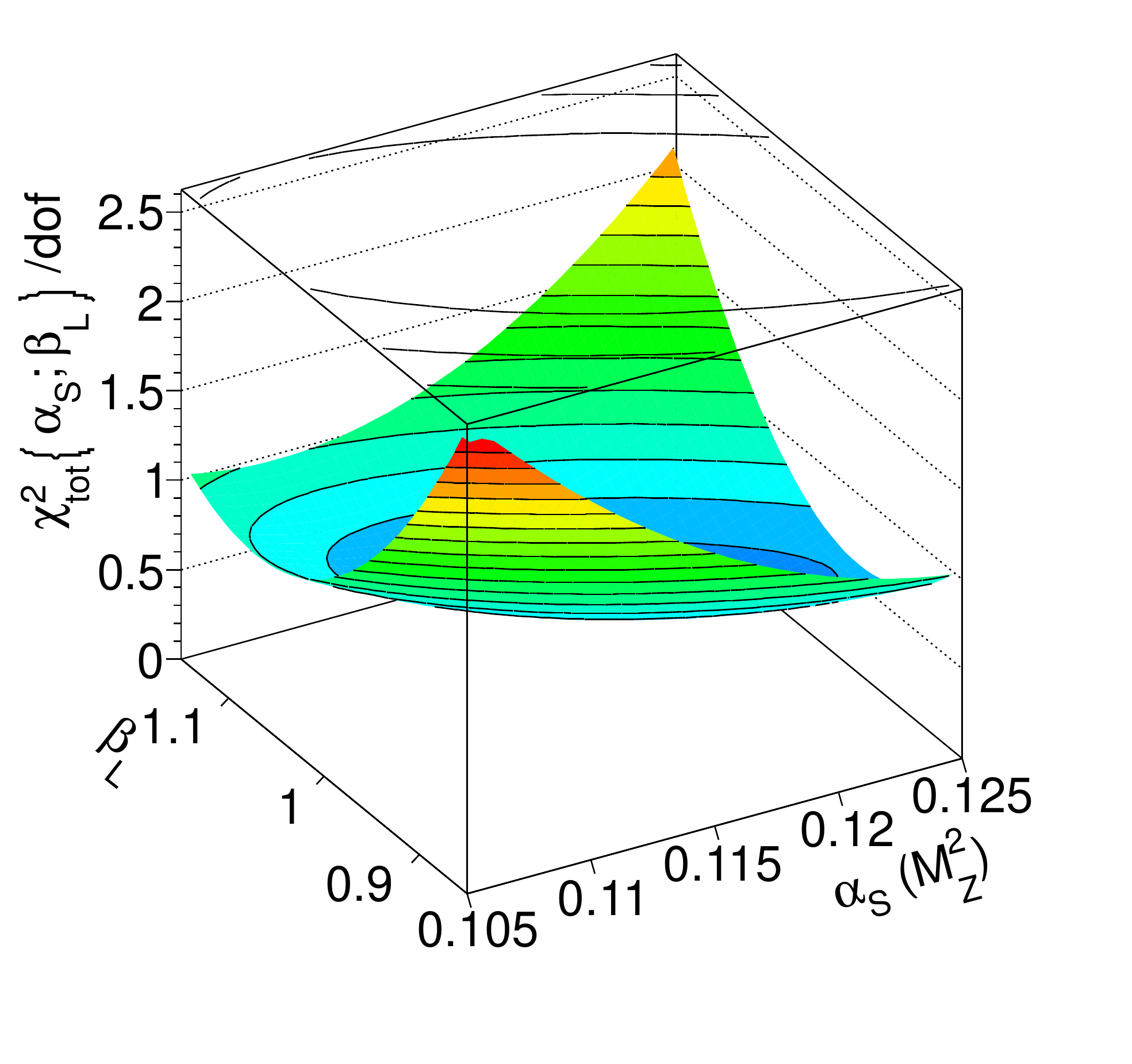}\hspace{0.4cm}
\includegraphics[width=8.cm]{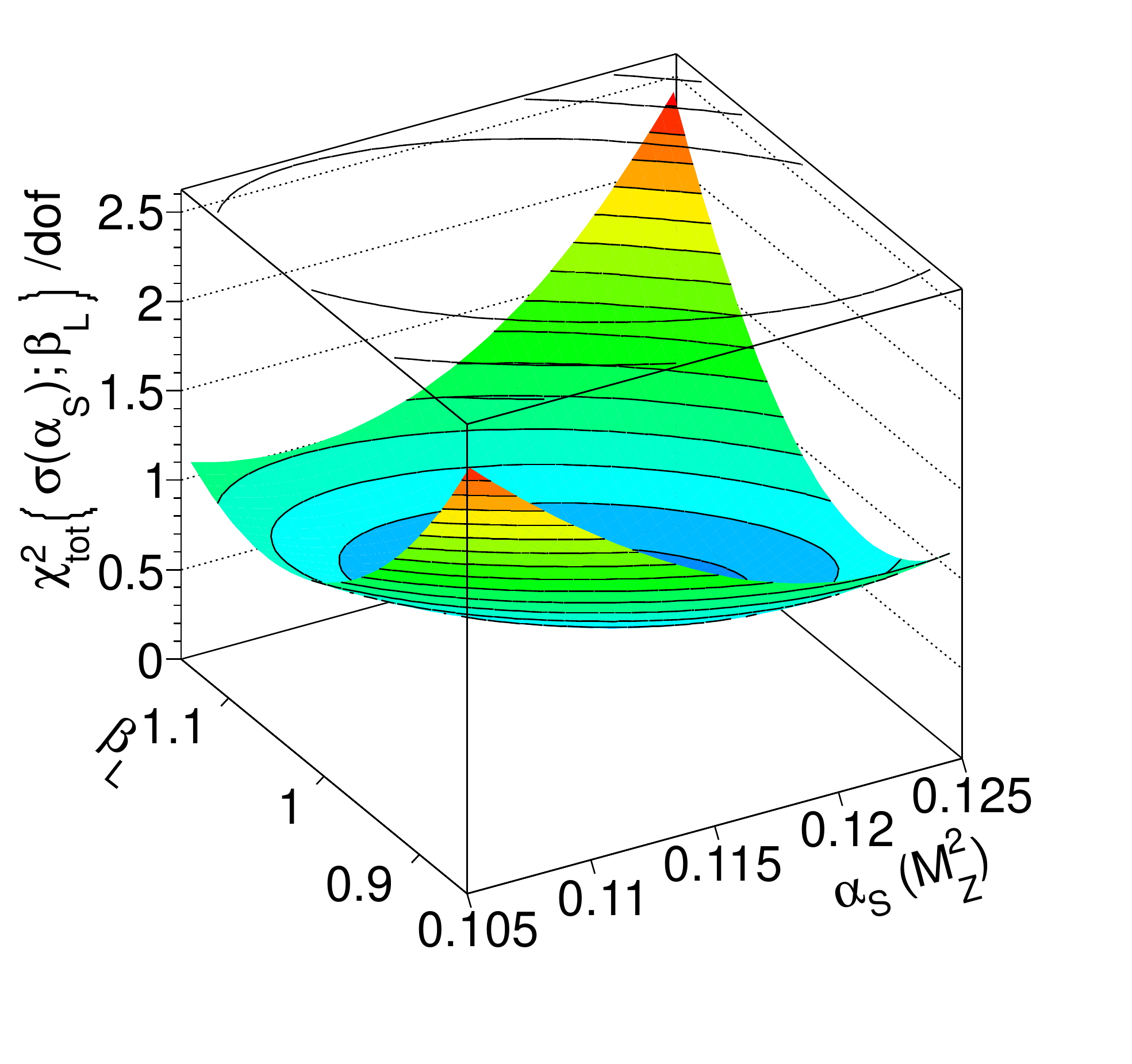}
\end{center}
\vspace{-0.5cm}
\caption[.]{ 
  $\chi^{2}/{\rm dof}$ values as a function of \asZ and $\beta_{\rm L}$~(see text).
  The top figures show the contribution from the comparison between data and theory, computed at the \as level~(left) and at the cross section level~(right).
  The center figure shows the contribution from the luminosity shifts, while the bottom figures show the total $\chi^{2}/{\rm dof}$, computed at the \as level~(left) and at the cross section level~(right).
}
\label{Fig:chi2ScansAsLumi}
\end{figure*}
Fig.~\ref{Fig:chi2ScansAsLumi} shows the result of the scans of the modified \chiSq values, as a function of \asZ and the luminosity shift parameter $\beta_{\rm L}$.
It shows the total \chiSq values, together with their sub-components, from the comparison between data and theory, as well as from the luminosity shift~(see section~\ref{SubSubSec:ModifiedChisq}).

The partial \chiSq from the comparison between data and theory, computed at the \as level, exhibits a degeneracy between \as and $\beta_{\rm L}$~(see Fig.~\ref{Fig:chi2ScansAsLumi}, top left).
This is expected since, for this partial \chiSq, a value obtained for a given $\asAv\left(\beta_{\rm L}\right)$ corresponds to a combination of \as and $\beta_{\rm L}$ values~(cf.
Eq.~\ref{Eq:EffectiveAsLumiDependence} which, for a fixed $\asAv\left(\beta_{\rm L}\right)$, admits an infinity of solutions).
The degeneracy is removed when adding the constraint on the luminosity shifts, which fixes the minimum of the global \chiSq at $\beta_{\rm L} = 1$.
Therefore, the minimal value of the global \chiSq is equal to the minimal value of the partial \chiSq from the comparison between data and theory, and it is obtained for $\asZ = 0.1175$.
One should recall, however, that the covariance matrix used here in the computation of the partial \chiSq from the comparison between data and theory is not complete, as the luminosity uncertainty
is treated as a separate nuisance parameter.
This explains the difference comparing to the result obtained from the minimisation of the ``standard'' global \chiSq, in the previous subsection.

The degeneracy between \as and $\beta_{\rm L}$, seen for the partial \chiSq from the comparison between data and theory, computed at the \as level, is absent for the partial \chiSq computed at
the cross section level~(see Fig.~\ref{Fig:chi2ScansAsLumi}, top right).
This is due to the slightly different dependence on \as, for the~(shapes of the) theoretical calculations in the various (\pt; \absy) bins~(the same effect which made
Eq.~\ref{Eq:EffectiveAsLumiDependence} to be valid only as an approximation).
While the above mentioned degeneracy is removed, this partial \chiSq does not allow for a strong constraint of the uncertainty on the luminosity.
After adding the constraint term and minimising the total \chiSq, the uncertainty on $\beta_{\rm L}$ is very similar to the input precision of the luminosity determination~($\approx 3.4\%$).
The minimum of this global \chiSq is reached by a shift of the nominal luminosity which is smaller than $0.5\%$~(much smaller than the uncertainty on the luminosity), and an \asZ value of $0.1160$.

While these two alternative \chiSq definitions still have some of the problems present for the ``standard'' global \chiSq~(in particular, the negative weights obtained in the minimisation
of the partial \chiSq from the comparison between data and theory), they show that the measurement of the inclusive jet cross section alone does not allow for an improved luminosity determination
when measuring \asZ.

\subsection{Results using a geometrical mean}

Using a geometrical average~(GA) for the combination, we obtain:
\beqns
   \as^{\rm GA} = 0.1149^{+0.0046}_{-0.0049}.
\eeqns
This value and its (slightly asymmetric)~uncertainties are very similar to the ones obtained using the simple and the weighted averages.
The same statement is true when comparing the corresponding averages in the various $|y|$ bins.

\subsection{Conclusions on the experimental results}

In the previous sections we have shown results on the \asZ determination, using several averaging procedures.
The methods using a matrix inversion in the \chiSq definition were observed to have badly behaved weights~(negative or larger than one) sometimes yielding an average outside the range of
the input values.
Alternative averaging methods, relying on the bin-to-bin correlations for each component of the systematic uncertainties, have been proposed in the past.
However, in our study, the only systematic uncertainty for which the correlations are perfectly known is the one on the ATLAS luminosity.
We have shown that our study does not allow for an improvement of the luminosity determination, while performing an \asZ measurement.

The results obtained using various weighted averages~(with simple weights, or using the inverse of squared uncertainties) were shown to have~(as expected) larger variances comparing to the ones
using the minimisation of a \chiSq with correlations.
However, these simpler weighted averages have well behaved weights, which do not rely on the exact information on the correlations, while all the information on the uncertainties and their
correlations is used in the error propagation~(through pseudo-experiments).
In addition, these results are confirmed by the ones obtained using a geometrical average.
It is for all these reasons that we choose to use the value obtained with the weights given by the inverse of the squared total uncertainties, for our nominal result.

In Fig.~\ref{Fig:asWeightedAvAllEta} we have shown the nominal result for \asZ, its uncertainty, and the comparison with the inputs used in the combination.
The general agreement is good~(except maybe for one bin from the forward region, where the theoretical prediction is more questionable), which is also confirmed by 
the global $\chiSq/{\rm dof}$~(see Sec.~\ref{Sec:ResultsWeightedAverage}).
\begin{figure*}[htbp]
\begin{center}
\includegraphics[width=8.7cm]{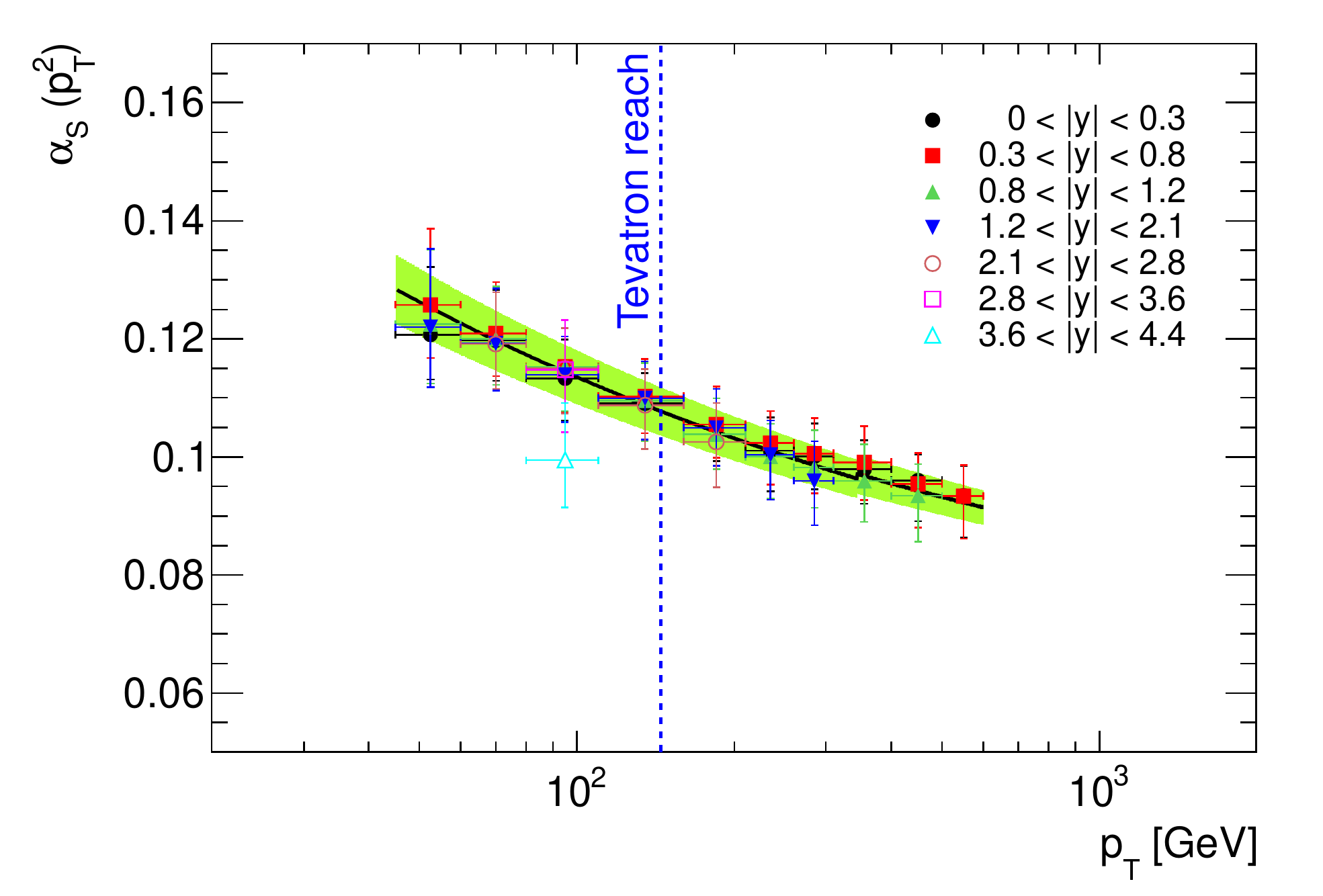}\hspace{0.4cm}
\includegraphics[width=8.7cm]{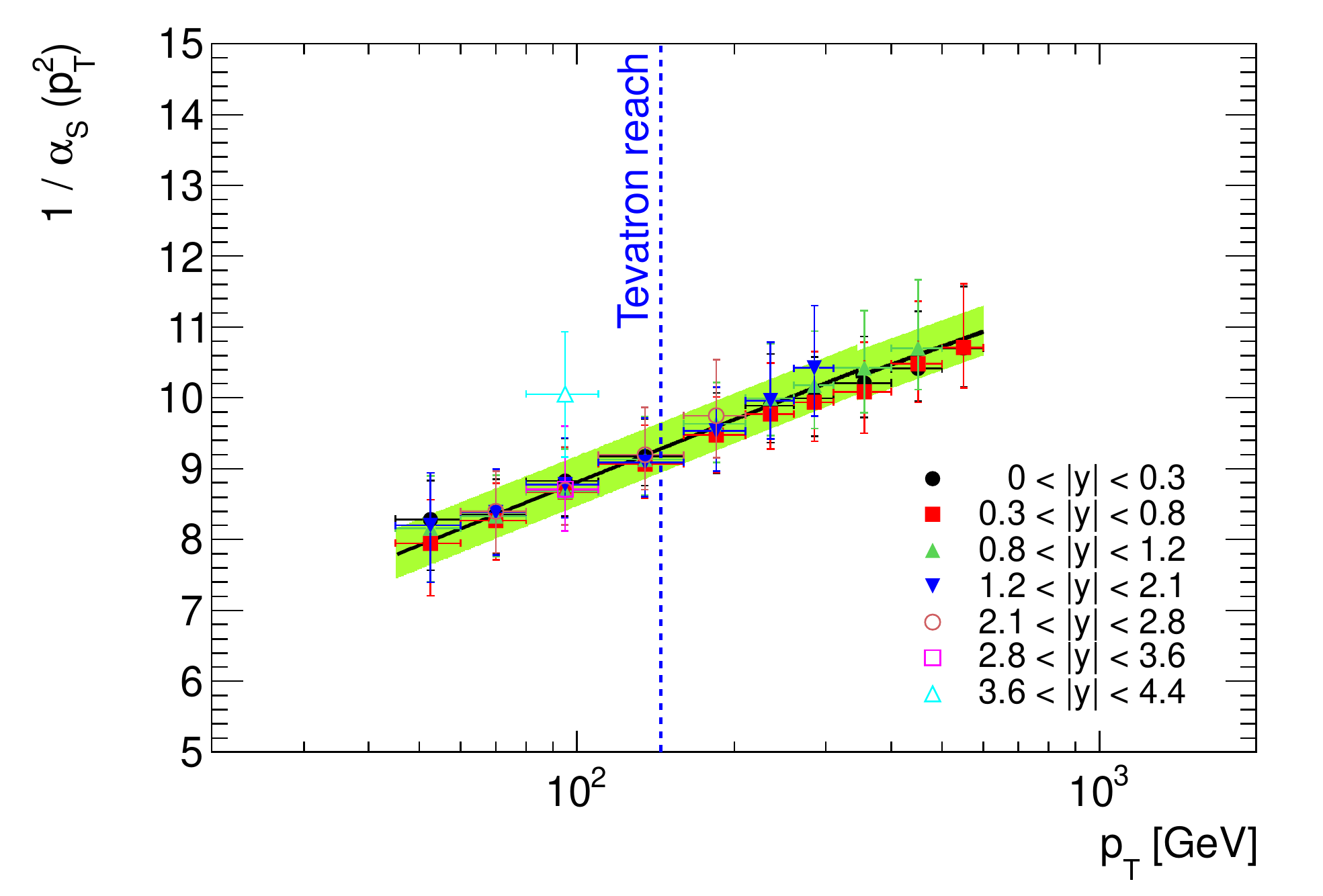}
\end{center}
\vspace{-0.5cm}
\caption[.]{ 
  Global weighted $\as$ average~(green band) evolved to the corresponding $p_{\rm T}$ scale~(left plot) and its inverse~(right plot), together with the values obtained in all the (\pt; \absy) bins
  used in the combination.
  The blue vertical line indicates the highest \pt value used in the Tevatron \asZ determination, from the inclusive jet cross section.
}
\label{Fig:asRunning}
\end{figure*}

The ${\rm N}^{3}{\rm LO}$ RGE running based on Runge-Kuta integration, with 3-loop quark-flavour matching at the top threshold~\cite{Chetyrkin:1997sg,Rodrigo:1997zd}, has been used
to evolve the \as values, from the Z scale to a given \pt scale.
This has been done for the global average, as well as for the values obtained in the individual (\pt; \absy) bins~(see Fig.~\ref{Fig:asRunning}).
The reduction of the relative uncertainty on the \as average, when performing the running from low to high \pt values, is also visible.
In the same figure, we indicate the highest \pt value~($145$~GeV) used in the \asZ determination, from the inclusive jet cross section, at Tevatron~\cite{Abazov:2009nc,Bandurin:2011sh}.
Our combination procedure uses a \pt range going up to four times higher.
We also use inputs from the forward region, up to $\absy = 4.4$, while the Tevatron determination is restricted to $\absy < 1.6$.

The SM running of \as is assumed in the various PDF determinations, as well as in the theoretical calculation of the inclusive jet cross section.
A possible contribution from ``New Physics'' would manifest by a deviation between the running of the \as average and the individual values.
This would probably be more ``visible'' in the figure showing the running of the inverse of \as~(see Fig.~\ref{Fig:asRunning})~\cite{Quigg}, although one should keep in mind that the same 
statistical information is available in all these various ways of presenting the same result.
A good agreement between our nominal fit and the $21$ data points in the newly explored region, above $145$~GeV, can be seen in Fig.~\ref{Fig:asRunning} and is also shown by
the corresponding partial \chiSq, accounting for $8.5$.
The weighted average computed in the same region alone yields $0.1157^{+0.0054}_{-0.0059}$, with a $\chiSq/{\rm dof}$ of $0.42$~(for 20 degrees of freedom).
This result is compatible with~(but slightly less precise than) the nominal weighted average computed for 42 data points~(Eq.~\ref{Eq:NominalWeightedAverage}).

As explained above, because of the limited range in the \as scan for the theoretical prediction, not all the (\pt; \absy) bins were used in the nominal combination.
This limited range in the \as scan induces distortions in the tails of the \as distributions for the corresponding bins.
The nominal values are, however, less affected by this problem.
Their combination allows for determining a systematic uncertainty due to the bins which where removed in the nominal combination, accounting for $\approx 0.0014$.

Results with a similar precision are obtained using \AKT jets with R=0.4.
The central value of the corresponding weighted average is, however, shifted downwards with respect to our nominal result~(from \AKT jets with R=0.6), by about $0.0060$.
This is expected, due to a systematic difference seen between the comparisons of the two (strongly correlated)~measurements with the corresponding theoretical predictions~(see
section~\ref{Sec:InputData}).
The study of the \chiSq scan~(see section~\ref{Sec:Chi2Scan}) applied to \AKT jets with R=0.4, indicates an even smaller ``preferred'' shift of the luminosity value, comparing to the one observed for
\AKT jets with R=0.6.
It should be pointed out that a shift in the nominal value of the luminosity~(or of any other nuisance parameter which is strongly correlated between the two measurements) could not explain
the observed difference anyway, as these shifts should be identical for \AKT jets with R=0.4 and R=0.6.
The difference between the \as determinations with the two jet sizes is treated as systematic uncertainty in our final result.
This is actually our largest uncertainty and an improvement in the understanding of these differences is desirable.
This improved understanding could be achieved, for example, through the measurement of the ratio between the spectra of \AKT jets with R=0.4 and R=0.6, and the comparison with the corresponding
theoretical prediction~(as given in~\cite{Soyez:2011np}).
The main challenge of this measurement is the understanding of the (strong)~correlations of the systematic uncertainties of the two spectra.

\subsection{Theoretical uncertainties}

The main uncertainties on the NLO QCD prediction come from the choice of the PDF set, the uncertainties on the nominal set of PDFs and the choice of factorisation and the renormalisation scales. 

In order to propagate the scale uncertainty on the cross section to the uncertainty on $\as$, the theoretical cross section was recalculated with six different scale choices
$
(\frac{\mu_R}{\mu_{R}^{\rm default}},\frac{\mu_F}{\mu_{F}^{\rm default}})
=\{(\frac{1}{2},1),\,(2,1),\,(1,\frac{1}{2}),\,(1,2),\,(\frac{1}{2},\frac{1}{2}),\,(2,2)\}
$
in each $(\pt,\;|y|)$ bin.
The nominal value of the measured cross section was mapped to the $\as\left(\mu_{F}, \mu_{R}\right)$ for each scale choice.
The asymmetric envelope of the values $\as\left(\mu_{F}, \mu_{R}\right)$ around the one with the default scale choice is taken as the scale uncertainty on $\as$. 
Propagating this uncertainty through our nominal averaging procedure results in a scale uncertainty on the \asZ value of $+0.0044\;-0.0011$.

For the evaluation of the uncertainty on $\as$ due to the choice of the PDF set, the theory prediction for the inclusive jet cross section was recalculated using MSTW~2008, NNPDF~2.1~(100)~\cite{Ball:2011mu, Ball:2011uy} and HERAPDF~1.5~\cite{HERAPDF15} PDF sets.
For each alternative PDF set the $\as^{PDF set}$ was determined from the measured cross section and the recalculated theoretical one, using the procedure described in Section~\ref{Sec:EvaluationProcedure}.
The estimated value of the PDF choice uncertainty is $+0.0022\;-0.0015$, which is the maximal difference between the $\as^{PDF set}$ and the central CT10~PDF one.

For the CT10~PDF the $N_{p}=26$ parameters~(eigenvectors) were determined. Therefore CT10 provides $2\times N_{p}$ PDF uncertainty sets for positive and negative variations of each eigenvector in the $N_{p}$-dimensional parameter space~\cite{Lai:2010vv}. The asymmetric PDF uncertainty on $\as$ for the CT10~PDF is given by~\cite{Campbell:2006wx}:
\begin{eqnarray*} 
\label{eqn:pdfuncertainty}
\Delta \as^{+} &=& \sqrt{\sum\limits_{i=1}^{N_{p}}\left[{\rm max}\left((\as^{+})_i - (\as)_0, (\as^{-})_i - (\as)_0, 0\right)\right]^2}\\
\Delta \as^{-} &=& \sqrt{\sum\limits_{i=1}^{N_{p}}\left[{\rm max}\left((\as)_0 - (\as)^{+}_i, (\as)_0 - (\as)^{-}_i, 0\right)\right]^2},
\end{eqnarray*}
where $(\as)_0$ and $(\as^{\pm})_i$ are calculated using the central PDF set and $\pm$ variation of all PDF set eigenvectors, respectively.
The PDF uncertainty on the average \asZ accounts for $\pm 0.0010$.

Finally, for the evaluation of the uncertainty on $\as$ due to the modelling of non-perturbative effects, the nominal value of the measured cross section was mapped
to the $\as^{MC_{tune}}$ for each MC tune used to estimate the  uncertainty of the non-perturbative corrections on the inclusive jet cross section.
Again, the asymmetric envelope of the values $\as^{MC_{tune}}$ around the one with the default tune is taken as the uncertainty on \asZ due to the modelling of non-perturbative effects,
yielding $+0.009\;-0.0034$.

Given the fact that the theoretical systematic uncertainties from the scale choice, PDF eigenvectors, the choice of the PDF set and modelling of non-perturbative effects,
are~(to a good approximation) independent, the total theoretical uncertainty is equal to their quadratic sum.


\section{Conclusions}
\label{Sec:Conclusions}

We have performed a determination of the strong coupling constant at the Z scale, using the ATLAS inclusive jet cross section data.
Our final result accounts for
\beqns
   \asZ &=& 0.1151  \pm 0.0001~({\rm stat.})  \pm 0.0047~({\rm exp.~syst.}) \\
   && \pm 0.0014~({\rm\pt~range})  \pm 0.0060~({\rm jet~size}) \\
   && ^{+ 0.0044}_{-0.0011}~({\rm scale}) ^{+0.0022}_{-0.0015}~({\rm PDF~choice})  \\
   && \pm 0.0010~({\rm PDF~eig.})  ^{+0.0009}_{-0.0034}~({\rm NP~corrections}) ,
\eeqns
where the uncertainties are statistical and experimental systematic~(propagated from the ATLAS data), due to the limited \pt range, 
due to the differences between the results obtained with the R=0.4 and R=0.6 jet sizes, due to the renormalisation and factorisation scale choice in the theoretical calculation, 
due to the choice of the PDF set, propagated from the PDF eigenvectors, and due to the non-perturbative corrections, respectively.
Our value is in good agreement with the latest (preliminary) update of the \asZ world average~($0.1183 \pm 0.0010$)~\cite{Bethke:2011tr}, as well as with the latest result from a hadron-hadron
collider~($0.1161^{+0.0041}_{-0.0048}$)~\cite{Abazov:2009nc,Bandurin:2011sh}.
Although our result is less precise, it includes for the first time the measurements of the inclusive jet cross section up to $600$~GeV.
The running of \as has also been tested, in the \pt range between $45$ and $600$~GeV, and no evidence of a deviation from the QCD prediction has been observed.

\begin{details}
~~ 
We thank T.~Carli for numerous discussions as well as a careful reading of this manuscript.
We acknowledge A.~Cooper-Sarkar, C.~Doglioni, A.~Glazov and A.~Hoecker for helpful discussions.
We benefited from the feedback of the ATLAS PDF forum and the ATLAS jet physics group on our study.
We gratefully acknowledge ATLAS Physics Coordination and ATLAS management for fruitful discussions and continuous support.
\end{details}

\end{document}